\documentstyle[12pt]{article}
\begin{document}
% Underline for text or math

\catcode`@=11
\def\un#1{\relax\ifmmode\@@underline#1\else
        $\@@underline{\hbox{#1}}$\relax\fi}
\catcode`@=12

% Accents and foreign (in text):

\let\under=\b                   % bar-under (but see \un above)
\let\ced=\c                     % cedilla
\let\du=\d                      % dot-under
\let\um=\H                      % Hungarian umlaut
\let\sll=\l                     % slashed (suppressed) l (Polish)
\let\Sll=\L                     % " L
\let\slo=\o                     % slashed o (Scandinavian)
\let\Slo=\O                     % " O
\let\tie=\t                     % tie-after (semicircle connecting two letters)
\let\br=\u                      % breve
                % Also: \`        grave
                %       \'        acute
                %       \v        hacek (check)
                %       \^        circumflex (hat)
                %       \~        tilde (squiggle)
                %       \=        macron (bar-over)
                %       \.        dot (over)
                %       \"        umlaut (dieresis)
                %       \aa \AA   A-with-circle (Scandinavian)
                %       \ae \AE   ligature (Latin & Scandinavian)
                %       \oe \OE   " (French)
                %       \ss       es-zet (German sharp s)
                %       \$  \#  \&  \%  \pounds  {\it\&}  \dots

% Abbreviations for Greek letters

\def\a{\alpha}
\def\b{\beta}
\def\c{\chi}
\def\d{\delta}
\def\e{\epsilon}
\def\f{\phi}
\def\g{\gamma}
\def\h{\eta}
\def\i{\iota}
\def\j{\psi}
\def\k{\kappa}
\def\l{\lambda}
\def\m{\mu}
\def\n{\nu}
\def\o{\omega}
\def\p{\pi}
\def\q{\theta}
\def\r{\rho}
\def\s{\sigma}
\def\t{\tau}
\def\u{\upsilon}
\def\x{\xi}
\def\z{\zeta}
\def\D{\Delta}
\def\F{\Phi}
\def\G{\Gamma}
\def\J{\Psi}
\def\L{\Lambda}
\def\O{\Omega}
\def\P{\Pi}
\def\Q{\Theta}
\def\S{\Sigma}
\def\U{\Upsilon}
\def\X{\Xi}

%Varletters
\def\ve{\varepsilon}
\def\vf{\varphi}
\def\vr{\varrho}
\def\vs{\varsigma}
\def\vq{\vartheta}

% Calligraphic letters

\def\ca{{\cal A}}
\def\cb{{\cal B}}
\def\cc{{\cal C}}
\def\cd{{\cal D}}
\def\ce{{\cal E}}
\def\cf{{\cal F}}
\def\cg{{\cal G}}
\def\ch{{\cal H}}
\def\ci{{\cal I}}
\def\cj{{\cal J}}
\def\ck{{\cal K}}
\def\cl{{\cal L}}
\def\cm{{\cal M}}
\def\cn{{\cal N}}
\def\co{{\cal O}}
\def\cp{{\cal P}}
\def\cq{{\cal Q}}
\def\car{{\cal R}}
\def\cs{{\cal S}}
\def\ct{{\cal T}}
\def\cu{{\cal U}}
\def\cv{{\cal V}}
\def\cw{{\cal W}}
\def\cx{{\cal X}}
\def\cy{{\cal Y}}
\def\cz{{\cal Z}}

% Fonts

\def\Sc#1{{\hbox{\sc #1}}}      % script for single characters in equations
\def\Sf#1{{\hbox{\sf #1}}}      % sans serif for single characters in equations

                        % Also:  \rm      Roman (default for text)
                        %        \bf      boldface
                        %        \it      italic
                        %        \mit     math italic (default for equations)
                        %        \sl      slanted
                        %        \em      emphatic
                        %        \tt      typewriter
                        % and sizes:    \tiny
                        %               \scriptsize
                        %               \footnotesize
                        %               \small
                        %               \normalsize
                        %               \large
                        %               \Large
                        %               \LARGE
                        %               \huge
                        %               \Huge

% Math symbols

\def\slpa{\slash{\pa}}                            % slashed partial derivative
\def\slin{\SLLash{\in}}                                   % slashed in-sign
\def\bo{{\raise-.5ex\hbox{\large$\Box$}}}               % D'Alembertian
\def\cbo{\Sc [}                                         % curly "
\def\pa{\partial}                                       % curly d
\def\de{\nabla}                                         % del
\def\dell{\bigtriangledown}                             % hi ho the dairy-o
\def\su{\sum}                                           % summation
\def\pr{\prod}                                          % product
\def\iff{\leftrightarrow}                               % <-->
\def\conj{{\hbox{\large *}}}                            % complex conjugate
\def\ltap{\raisebox{-.4ex}{\rlap{$\sim$}} \raisebox{.4ex}{$<$}}   % < or ~
\def\gtap{\raisebox{-.4ex}{\rlap{$\sim$}} \raisebox{.4ex}{$>$}}   % > or ~
\def\TH{{\raise.2ex\hbox{$\displaystyle \bigodot$}\mskip-4.7mu \llap H \;}}
\def\face{{\raise.2ex\hbox{$\displaystyle \bigodot$}\mskip-2.2mu \llap {$\ddot
        \smile$}}}                                      % happy face
\def\dg{\sp\dagger}                                   % hermitian conjugate
\def\ddg{\sp\ddagger}                                   % double dagger
\def\di{\int\!\!\!\int}                                 % double integral
                        % Also:  \int  \oint              integral, contour
                        %        \hbar                    h bar
                        %        \infty                   infinity
                        %        \sqrt                    square root
                        %        \pm  \mp                 plus or minus
                        %        \cdot  \cdots            centered dot(s)
                        %        \oplus  \otimes          group theory
                        %        \equiv                   equivalence
                        %        \sim                     ~
                        %        \approx                  approximately =
                        %        \proptoto                  funny alpha
                        %        \ne                      not =
                        %        \le \ge                  < or = , > or =
                        %        \{  \}                   braces
                        %        \to  \gets               -> , <-
                        % and spaces:  \,  \:  \;  \quad  \qquad
                        %              \!                 (negative)

\font\tenex=cmex10 scaled 1200

% Math stuff with one argument

\def\sp#1{{}^{#1}}                              % superscript (unaligned)
\def\sb#1{{}_{#1}}                              % sub"
\def\oldsl#1{\rlap/#1}                          % poor slash
\def\slash#1{\rlap{\hbox{$\mskip 1 mu /$}}#1}      % good slash for lower case
\def\Slash#1{\rlap{\hbox{$\mskip 3 mu /$}}#1}      % " upper
\def\SLash#1{\rlap{\hbox{$\mskip 4.5 mu /$}}#1}    % " fat stuff (e.g., M)
\def\SLLash#1{\rlap{\hbox{$\mskip 6 mu /$}}#1}      % slash for no-in sign
\def\PMMM#1{\rlap{\hbox{$\mskip 2 mu | $}}#1}   %
\def\PMM#1{\rlap{\hbox{$\mskip 4 mu ~ \mid $}}#1}       %
\def\Tilde#1{\widetilde{#1}}                    % big tilde
\def\Hat#1{\widehat{#1}}                        % big hat
\def\Bar#1{\overline{#1}}                       % big bar
\def\bra#1{\left\langle #1\right|}              % < |
\def\ket#1{\left| #1\right\rangle}              % | >
\def\VEV#1{\left\langle #1\right\rangle}        % < >
\def\abs#1{\left| #1\right|}                    % | |
\def\leftrightarrowfill{$\mathsurround=0pt \mathord\leftarrow \mkern-6mu
        \cleaders\hbox{$\mkern-2mu \mathord- \mkern-2mu$}\hfill
        \mkern-6mu \mathord\rightarrow$}
\def\dvec#1{\vbox{\ialign{##\crcr
        \leftrightarrowfill\crcr\noalign{\kern-1pt\nointerlineskip}
        $\hfil\displaystyle{#1}\hfil$\crcr}}}           % <--> accent
% Next line previously contained {\LARGE .}
\def\dt#1{{\buildrel {\hbox{.}} \over {#1}}}     % dot-over for sp/sb
\def\dtt#1{{\buildrel \bullet \over {#1}}}              % alternate "
\def\der#1{{\pa \over \pa {#1}}}                % partial derivative
\def\fder#1{{\d \over \d {#1}}}                 % functional derivative
                % Also math accents:    \bar
                %                       \check
                %                       \hat
                %                       \tilde
                %                       \acute
                %                       \grave
                %                       \breve
                %                       \dot    (over)
                %                       \ddot   (umlaut)
                %                       \vec    (vector)

% Math stuff with more than one argument

\def\frac#1#2{{\textstyle{#1\over\vphantom2\smash{\raise.20ex
        \hbox{$\scriptstyle{#2}$}}}}}                   % fraction
\def\ha{\frac12}                                        % 1/2
\def\sfrac#1#2{{\vphantom1\smash{\lower.5ex\hbox{\small$#1$}}\over
        \vphantom1\smash{\raise.4ex\hbox{\small$#2$}}}} % alternate fraction
\def\bfrac#1#2{{\vphantom1\smash{\lower.5ex\hbox{$#1$}}\over
        \vphantom1\smash{\raise.3ex\hbox{$#2$}}}}       % "
\def\afrac#1#2{{\vphantom1\smash{\lower.5ex\hbox{$#1$}}\over#2}}    % "
\def\partder#1#2{{\partial #1\over\partial #2}}   % partial derivative of
\def\parvar#1#2{{\d #1\over \d #2}}               % variation of
\def\secder#1#2#3{{\partial^2 #1\over\partial #2 \partial #3}}  % second "
\def\on#1#2{\mathop{\null#2}\limits^{#1}}               % arbitrary accent
\def\bvec#1{\on\leftarrow{#1}}                  % backward vector accent
\def\oover#1{\on\circ{#1}}                              % circle accent

\def\[{\lfloor{\hskip 0.35pt}\!\!\!\lceil}
\def\]{\rfloor{\hskip 0.35pt}\!\!\!\rceil}
\def\Lag{{\cal L}}
\def\du#1#2{_{#1}{}^{#2}}
\def\ud#1#2{^{#1}{}_{#2}}
\def\dud#1#2#3{_{#1}{}^{#2}{}_{#3}}
\def\udu#1#2#3{^{#1}{}_{#2}{}^{#3}}
\def\calD{{\cal D}}
\def\calM{{\cal M}}

\def\szet{{${\scriptstyle \b}$}}
\def\ulA{{\un A}}
\def\ulM{{\underline M}}
\def\cdm{{\Sc D}_{--}}
\def\cdp{{\Sc D}_{++}}
\def\vTheta{\check\Theta}
\def\gg{{\hbox{\sc g}}}
\def\fracm#1#2{\hbox{\large{${\frac{{#1}}{{#2}}}$}}}
\def\half{{\fracm12}}
\def\ha{\half}
\def\tr{{\rm tr}}
\def\Tr{{\rm Tr}}
\def\itrema{$\ddot{\scriptstyle 1}$}
\def\ula{{\underline a}} \def\ulb{{\underline b}} \def\ulc{{\underline c}}
\def\uld{{\underline d}} \def\ule{{\underline e}} \def\ulf{{\underline f}}
\def\ulg{{\underline g}}
\def\items#1{\\ \item{[#1]}}
\def\ul{\underline}
\def\un{\underline}
\def\fracmm#1#2{{{#1}\over{#2}}}
\def\footnotew#1{\footnote{\hsize=6.5in {#1}}}
\def\low#1{{\raise -3pt\hbox{${\hskip 0.75pt}\!_{#1}$}}}

\def\Dot#1{\buildrel{_{_{\hskip 0.01in}\bullet}}\over{#1}}
\def\dt#1{\Dot{#1}}
\def\DDot#1{\buildrel{_{_{\hskip 0.01in}\bullet\bullet}}\over{#1}}
\def\ddt#1{\DDot{#1}}

%\def\Tilde#1{{\widetilde{#1}}\hskip 0.015in}
%\def\Hat#1{\widehat{#1}}

% Aligned equations

\newskip\humongous \humongous=0pt plus 1000pt minus 1000pt
\def\caja{\mathsurround=0pt}
\def\eqalign#1{\,\vcenter{\openup2\jot \caja
        \ialign{\strut \hfil$\displaystyle{##}$&$
        \displaystyle{{}##}$\hfil\crcr#1\crcr}}\,}
\newif\ifdtup
\def\panorama{\global\dtuptrue \openup2\jot \caja
        \everycr{\noalign{\ifdtup \global\dtupfalse
        \vskip-\lineskiplimit \vskip\normallineskiplimit
        \else \penalty\interdisplaylinepenalty \fi}}}
\def\li#1{\panorama \tabskip=\humongous                         % eqalignno
        \halign to\displaywidth{\hfil$\displaystyle{##}$
        \tabskip=0pt&$\displaystyle{{}##}$\hfil
        \tabskip=\humongous&\llap{$##$}\tabskip=0pt
        \crcr#1\crcr}}
\def\eqalignnotwo#1{\panorama \tabskip=\humongous
        \halign to\displaywidth{\hfil$\displaystyle{##}$
        \tabskip=0pt&$\displaystyle{{}##}$
        \tabskip=0pt&$\displaystyle{{}##}$\hfil
        \tabskip=\humongous&\llap{$##$}\tabskip=0pt
        \crcr#1\crcr}}

% Journal abbreviations (preprints)

\def\NPB{{\sf Nucl. Phys. }{\bf B}}
\def\PL{{\sf Phys. Lett. }}
\def\PRL{{\sf Phys. Rev. Lett. }}
\def\PRD{{\sf Phys. Rev. }{\bf D}}
\def\CQG{{\sf Class. Quantum Grav. }}
\def\JMP{{\sf J. Math. Phys. }}
\def\SJNP{{\sf Sov. J. Nucl. Phys. }}
\def\SPJ{{\sf Sov. Phys. J. }}
\def\JETPL{{\sf JETP Lett. }}
\def\TMP{{\sf Theor. Math. Phys. }}
\def\IJMPA{{\sf Int. J. Mod. Phys. }{\bf A}}
\def\MPL{{\sf Mod. Phys. Lett. }}
\def\CMP{{\sf Commun. Math. Phys. }}
\def\AP{{\sf Ann. Phys. }}
\def\PR{{\sf Phys. Rep. }}

\def\app#1#2#3{Acta Phys.~Pol.~{\bf B{#1}} (19{#2}) #3}
\def\pl#1#2#3{Phys.~Lett.~{\bf {#1}B} (19{#2}) #3}
\def\np#1#2#3{Nucl.~Phys.~{\bf B{#1}} (19{#2}) #3}
\def\prl#1#2#3{Phys.~Rev.~Lett.~{\bf #1} (19{#2}) #3}
\def\pr#1#2#3{Phys.~Rev.~{\bf D{#1}} (19{#2}) #3}
\def\cqg#1#2#3{Class.~and Quantum Grav.~{\bf {#1}} (19{#2}) #3}
\def\cmp#1#2#3{Commun.~Math.~Phys.~{\bf {#1}} (19{#2}) #3}
\def\jmp#1#2#3{J.~Math.~Phys.~{\bf {#1}} (19{#2}) #3}
\def\ap#1#2#3{Ann.~Phys.~(NY)~{\bf {#1}} (19{#2}) #3}
\def\prep#1#2#3{Phys.~Rep.~{\bf {#1}C} (19{#2}) #3}
\def\ptp#1#2#3{Progr.~Theor.~Phys.~{\bf {#1}} (19{#2}) #3}
\def\ijmp#1#2#3{Int.~J.~Mod.~Phys.~{\bf A{#1}} (19{#2}) #3}
\def\mpl#1#2#3{Mod.~Phys.~Lett.~{\bf A{#1}} (19{#2}) #3}
\def\nc#1#2#3{Nuovo Cim.~{\bf {#1}} (19{#2}) #3}
\def\ibid#1#2#3{{\it ibid.}~{\bf {#1}} (19{#2}) #3}
\def\sjnp#1#2#3{Sov.~J.~Nucl.~Phys.{\bf {#1}} (19{#2}) #3}
\def\tmp#1#2#3{Theor.~Math.~Phys.{\bf {#1}} (19{#2}) #3}
% Text style parameters

\topmargin=0in                          % top margin (less 1") (LaTeX)
\headheight=0in                         % height of heading (LaTeX)
\headsep=0in                    % separation of heading from body (LaTeX)
\textheight=9in                         % height of body (LaTeX)
%\footheight=3ex                         % height of foot (LaTeX)
\footskip=4ex           % distance between bottoms of body & foot (LaTeX)
\textwidth=6in                          % width of body (LaTeX)
\hsize=6in                              % " (TeX)
\parskip=\medskipamount                 % space between paragraphs (LaTeX)
\lineskip=0pt                           % minimum box separation (TeX)
\abovedisplayskip=1em plus.3em minus.5em        % space above equation (either)
\belowdisplayskip=1em plus.3em minus.5em        % " below
\abovedisplayshortskip=.5em plus.2em minus.4em  % " above when no overlap
\belowdisplayshortskip=.5em plus.2em minus.4em  % " below
\def\baselinestretch{1.2}       % magnification for line spacing (LaTeX)
\thicklines                         % thick straight lines for pictures (LaTeX)

% Section heading and reference stuff

\def\sect#1{\bigskip\medskip \goodbreak \noindent{\bf {#1}} \nobreak \medskip}
\def\refs{\sect{References} \footnotesize \frenchspacing \parskip=0pt}
\def\Item{\par\hang\textindent}
\def\Itemitem{\par\indent \hangindent2\parindent \textindent}
\def\makelabel#1{\hfil #1}
\def\topic{\par\noindent \hangafter1 \hangindent20pt}
\def\Topic{\par\noindent \hangafter1 \hangindent60pt}

\renewcommand\cd{\!\cdot\!}
\renewcommand\={\,=\,}
\newcommand\ba{\Bar}
\newcommand\ra{\rightarrow}
\renewcommand\do{\!\sp\dagger}
\newcommand\el{\ell}
\renewcommand\tr{\,{\rm tr}}
\newcommand\sa{{\sf a}}
\newcommand\A{{\sf A}}
\newcommand\B{{\sf B}}
\newcommand\C{{\sf C}}
\newcommand\E{{\sf E}}
\renewcommand\F{{\sf F}}
\newcommand\Fp{\F\pr}
\renewcommand\G{{\sf G}}
\renewcommand\H{{\sf H}}
\newcommand\sg{{\sf g}}
\newcommand\I{{\sf I}}
\newcommand\K{{\sf K}}
\newcommand\V{{\sf V}}
\newcommand\hX{\Hat X}
\renewcommand\X{{\sf X}}
\newcommand\HX{\Hat\X}
\newcommand\Y{{\sf Y}}
\newcommand\Z{{\sf Z}}
\newcommand\ZM{{Z^{\scriptstyle M}}}
\newcommand\Zm{{\Z^{\scriptstyle M}}}
\newcommand\ZS{{Z^{\scriptstyle S}}}
\newcommand\Zs{{\Z^{\scriptstyle S}}}
\newcommand\HZ{\hat\Z}
\renewcommand\pr{\sp\prime}
\newcommand\ov{\over}
\renewcommand\ch{\choose}
\renewcommand\tt{{\tilde{\tr}}}
\newcommand\bee{\begin{equation}}
\newcommand\ene{\end{equation}}
\newcommand\bea{\begin{eqnarray}}
\newcommand\ena{\end{eqnarray}}
\newcommand\non{\nonumber}

\renewcommand{\theequation}{\arabic{section}.\arabic{equation}}

\thispagestyle{empty}
\hbox to\hsize{\vbox{\noindent DESY 97--162 \hfill August 1997}}
\noindent
\vskip1.8cm
\begin{center}
{\LARGE\bf Index-free Heat Kernel Coefficients}\\
\vglue.3in
Anton~E.~M. van de Ven
\vglue.3in
{\it II Institute for Theoretical Physics, University of Hamburg}\\
{\it Luruper Chaussee 149, 22761 Hamburg, Germany.}\\
{\it e-mail: vandeven@vxdesy.desy.de}
\vglue.6in
{\bf Abstract}
\end{center}
\noindent
Using index-free notation, we present the diagonal values $a_j(x,x)$ of the 
first five heat kernel coefficients $a_j(x,x')$ associated with a general 
Laplace-type operator on a compact Riemannian space without boundary. The 
fifth coefficient $a_5(x,x)$ appears here for the first time. For the special 
case of a flat space, but with a gauge connection, the sixth coefficient is 
given too. 
Also provided are the leading terms for any coefficient, both in ascending and
descending powers of the Yang-Mills and Riemann curvatures, to the same order 
as required for the fourth coefficient. These results are obtained by directly
solving the relevant recursion relations, working in Fock-Schwinger gauge and 
Riemann normal coordinates. Our procedure is thus noncovariant, but we show 
that for any coefficient the `gauged' respectively `curved' version is found 
from the corresponding `non-gauged' respectively `flat' coefficient by making 
some simple covariant substitutions. These substitutions being understood, the
coefficients retain their `flat' form and size. In this sense the fifth and 
sixth coefficient have only 26 and 75 terms respectively, allowing us to 
write them down. Using index-free notation also clarifies the general 
structure of the heat kernel coefficients. In particular, in flat space we
find that from the fifth coefficient onward, certain scalars are absent. 
This may be relevant for the anomalies of quantum field theories in ten or 
more dimensions.
\newpage
\section{Introduction}
The heat kernel method has become a ubiquitous tool in both mathematics and 
physics (see \cite{Fu1} for a recent overview). In mathematics it appears e.g.
in the study of the spectral geometry of a Laplace-type differential operator 
on a Riemannian space and in the proof of index theorems \cite{ABP,Gi1}. In 
physics, the euclidean one-loop effective action for a given quantum field 
theory can be expressed in terms of the determinant of such a differential 
operator \cite{DeW}, which in turn can be written in terms of the associated 
heat kernel. The heat kernel therefore appears in many places, from quantum 
gravity \cite{DeW} to chiral perturbation theory \cite{Bal}. Anomalies can 
also be studied with the heat kernel method (see \cite{Bal,Ber}). Physicists 
frequently refer to it as the Schwinger-DeWitt \cite{Sch,DeW} or proper-time 
method. Exact expressions for the heat kernel exist only for special spaces. 
In the general case one may use its asymptotic expansion in the 
proper-time\footnote{An alternative is the so-called covariant perturbation 
theory \cite{BarV}. It provides a (partial) summation of the Schwinger-DeWitt 
series and can account for nonlocal effects.}.
The coefficients in this expansion are the so-called heat kernel 
coefficients (see sect 2 for a precise definition).

Several methods have been developed to find the heat kernel coefficients (see
 e.g. \cite{AvS} for a review). DeWitt \cite{DeW} determined the first two 
coefficients\footnote{We frequently abbreviate `diagonal value of a heat 
kernel coefficient' to `coefficient'.} with a covariant recursive method.
Sakai \cite{Sak} relied on Riemannian coordinates to find the third coefficient
in the scalar case (i.e. a single scalar field on a curved space).
For the general case, this coefficient was found by Gilkey \cite{Gi2} using a 
noncovariant pseudo-differential-operator technique. The integrated and traced
fourth and fifth coefficients for an arbitrary field theory in flat space were 
found in \cite{vdV} through the evaluation of a noncovariant Feynman graph. 
Avramidi \cite{Avr} presented a new covariant nonrecursive procedure and found
the fourth coefficient for the general case (for the scalar case see also
 \cite{ABC}). More recently, string-inspired world line path integral methods 
have been used \cite{FHSS1} to determine the functional trace of the first 
eight heat kernel coefficients for the case of a matrix potential in flat 
space without gauge connection (see also \cite{BELS}). 

In this paper the explicit diagonal value of the fifth heat kernel coefficient
in the general case is presented for the first time. In physics this 
coefficient is of importance e.g. in analysing the short distance behavior and 
anomalies of ten-dimensional quantum field theories (see \cite{FrT}). 
However, the number of terms in the higher heat kernel coefficients grows 
rapidly, leading one to expect more than a thousand terms for the fifth 
coefficient\footnote{Even restricting to the a single scalar field on a curved
space, the $j$-th coefficient already contains $1,4,17,92,668$ $R^j$-terms for 
$j=1,2,3,4,5$ (see appendix A of \cite{FKW}).}. This would seem to preclude 
writing down this coefficient in an intelligable form. Indeed, a computer 
would appear to be an essential piece of equipment in determining and storing 
the fifth coefficient. Contrary to these expectations, we will show here that 
with a suitable index-free notation one obtains a compact expression for this 
coefficient, containing only 26 terms. 

Using standard matrix notation for the field indices, the heat kernel 
coefficients are scalars and this suggests that it may be possible to write 
them in a form which is free of spacetime indices as well. 
Since the heat kernel coefficients are covariant, we may determine them in a 
special gauge and adapted coordinates. Following earlier authors, we select
the Fock-Schwinger gauge and Riemann normal coordinates. 
Examination of the relevant recursion relations then shows that only certain 
maximally symmetrized multiple (covariant) derivatives of the matrix potential
and Yang-Mills or Riemann curvature tensors appear. The heat kernel 
coefficients being formal scalars, it
turns out that we need to keep track of the rank of these tensors only (note  
the similarity with totally antisymmetric tensors, i.e. differential forms). 
This provides the basis for our index-free notation. Using instead a fully 
covariant method, Avramidi \cite{Avr} has arrived at similar conclusions by 
expanding in so-called covariant Taylor series, obtaining in this way the 
first four coefficients. 
Our non-covariant procedure seems to be no less efficient and, by maintaining 
index-free notation and manifest hermiticity, yields even compacter answers for
the heat kernel coefficients. Thus, in the special case of a flat space with a
gauge connection, we can also present the answer for the sixth coefficient. It
contains only 75 terms.

Using index-free notation also permits us to investigate the general structure 
of the heat kernel coefficients. Without actually solving the recursion 
relations, we can show that certain Lorentz scalars are absent from all 
coefficients. In flat space, this happens for the first time in the fifth 
coefficient. This may therefore be relevant for the anomalies of quantum field
theories in ten or more dimensions.

As we already mentioned above, a brute force approach with a computer algebra
program would have produced an unwieldy result.
Our results were obtained without the aid of a computer. However, it is 
relatively easy to program our index-free method and we used FORM \cite{Ver} 
and Mathematica \cite{Wol} to run some checks.

An outline of this paper is as follows. In sect 2 we recall the main features
of the heat kernel method {\it alias} the Schwinger-DeWitt formalism. In sect 
3 we introduce our index-free notation and use it to determine first the heat 
kernel coefficients in flat space without gauge connection, expanding in powers
of either derivatives (sect 3.1) or of the matrix potential (sect 3.2). 
In sect 3.1 we also use our index-free notation to prove that certain scalars 
are absent from all heat kernel coefficients. In sect 4 we show that the 
corresponding heat kernel coefficients with a gauge connection can be obtained
from a simple covariantization process. This involves not only replacing 
partial derivatives by covariant ones but also adding new field strength 
dependent terms. The latter kind of terms are shown to arise only as shifts in
the potential and in its covariant derivatives. These shifts being understood,
the heat kernel coefficients do not change their form upon `turning on the 
gauge field'. In sect 5 we employ Riemann normal coordinates to generalize to 
a curved space. In particular, we present in subsect 5.1 an explicit expansion
of the vielbein to all orders in these coordinates (such an expansion, usually
given for the metric, but only to some finite order in the normal coordinates 
can be found in many places, e.g. \cite{Sak,McL}). We give a similar result in
subsect 5.2 for the gauge connection to all orders in normal coordinates. 
Based on this, we find in subsect 5.3 the explicit form for any heat kernel 
coefficient up to and including terms of fourth order in the Yang-Mills and 
Riemann curvatures. The curved coefficients can be obtained from the 
corresponding gauged but flat coefficients via further simple covariant 
substitutions. To complete the fifth coefficient, we need to find the few 
terms of fifth order in the curvatures. This we do in subsect 5.4 by 
specializing to a locally symmetric space. 
In sect 6 we present the explicit answers for the first five coefficients. 
We indicate how to return to more conventional notation and compare with 
earlier results. Our conclusions are given in sect 7. Several appendices 
follow (in particular, the sixth coefficient in flat space is given in 
appendix C).
\section{Schwinger-DeWitt formalism}
Consider a set of fields $\f_i(x)$, $i=1\dots n$, defined over a compact
$d$-dimensional Riemannian manifold\footnote{The fields can be considered 
to be sections of a smooth vector bundle. The heat kernel coefficients do 
not explicitly depend on the dimension or signature of space(time). We assume
space to have no boundary (for the case with boundary, see e.g. \cite{BGV}).}
with coordinates $x^\m$, $\m=1\dots d$ and metric $g_{\m\n}$ (see appendix A 
for our notation and conventions). The fields are acted upon by a Laplace-type
wave operator $\D$
\bee\label{wave}
\D \= -\, \de^2 - X \quad ,\quad \de^2 \= g^{\m\n}\de_\m\de_\n
\ene
Here the covariant derivative $\de$ includes connection terms as needed for 
the fields $\f_i$. $X$ is a hermitian $n\times n$ matrix potential (we suppress
the field or `bundle' indices). The wave operator $\D$ is hermitian with 
respect to the inner product
\bee
(\f ,\j) \= \int d^dx \sqrt{g}\ \f^\ast \j
\ene
For most bosonic gauge field theories of interest one can achieve a wave 
operator of Laplace-type as in (\ref{wave}) by a suitable gauge choice. For 
fermionic (gauge) fields one squares the wave operator to obtain again 
(\ref{wave}) (see \cite{BaV} for wave operators not of this form).

Following Schwinger and DeWitt, we introduce the proper-time parameter $\t$
and define the heat kernel $K$ associated with $\D$ by
\bee\label{hkd}
({\pa\ov\pa\t} +\D)\, K(x,x\pr;\t) \= 0 \quad ,\quad
K(x,x\pr;0) \= I\d(x,x')
\ene
where $I$ is the $n\times n$ unit matrix and the bi-scalar $\d$ function is
defined by
\bee
\int d^dx\, {\sqrt g}\,\d(x,x')\,\f(x) \= \f(x\pr) 
\ene
for any scalar field $\f$. As we mentioned in the introduction, an exact 
solution for the kernel $K$ exists only for special spaces. We instead make
DeWitt's ansatz
\bee\label{hkc}
K(x,x\pr;\t) \= (4\p\t)^{-d/2} {\cal D}(x,x\pr)^{1/2} e^{-\s(x,x\pr)/2\t}
\su_{j=0}^\infty  a_j(x,x\pr)\,{\t^j\ov j!} 
\ene
which is known to be an asymptotic expansion in $\t$ \cite{Gi2}. Note our
unconventional normalization for the heat kernel coefficients $a_j$, which 
however agrees with \cite{Avr}. They transform as scalar densities of weight 
$-1/2$ at both $x$ and $x\pr$. The bi-scalar $\s$ is the geodetic interval 
(one half of the distance squared between $x$ and $x\pr$) and satisfies
\bee\label{sig}
\s_;\sp\m\s_{;\m} \= 2\s\quad ,\quad [\s] \equiv\s(x,x) \= 0
\ene
where we use Synge's bracket notation to indicate evaluation on the diagonal. 
The bi-scalar ${\cal D}$ is the Van Vleck-Morette determinant defined by
\bee
{\cal D}(x,x\pr) \= 
g^{-1/2}\,\det(-\s_{;\m\n\pr}) \, {g\pr}{\phantom g\!\!}^{-1/2}
\ene
where a prime refers to the point $x\pr$. It satisfies
\bee\label{vvm}
(2\s_;\sp\m\de_\m + \s_;\sp\m\sb\m - d)\, {\cal D}^{1/2} \= 0
\quad ,\quad [{\cal D}] \= 1
\ene
Inserting (\ref{hkc}) into (\ref{hkd}) and using (\ref{sig}) and (\ref{vvm}), 
one finds that the heat kernel coefficients must satisfy the following 
recursion relations for $j\geq 0$
\bee\label{ajd}
(\s^\m\de_\m + j) a_j \= -\, j {\cal D}^{-1/2} \D\, {\cal D}^{1/2} a_{j-1}
\quad ,\quad [a_0] \= I
\ene
where it is to be understood that $a_{-1}$ vanishes. Note that whereas $\s$ 
and ${\cal D}$ depend only on the metric, the $a_j$ are matrix valued and 
depend in addition on the detailed form of the wave operator. The hermiticity 
of the wave operator implies
\bee
a_j(x,x\pr)^\dagger \= a_j(x\pr,x) \quad\Rightarrow\quad [a_j]^\dagger\= [a_j]
\ene
To keep this property manifest we introduce the following notation: for any 
matrix valued function $F$ we define
\bee\label{brac}
\{F\}\equiv F\, +\, F^\dagger\quad {\rm if}\quad F\neq F^\dagger\quad 
{\rm else}\quad F
\ene
Thus $\{F\}=F$ when $F$ is selfadjoint. Frequently one is interested in the 
functional trace of the heat kernel coefficients
\bee
b_j \,\equiv\, {\rm Tr} \, a_j\equiv \tr\,\int d^dx\,\sqrt{g}\ a_j(x,x)
\ene
where $\tr$ denotes the matrix trace over field indices only. To determine e.g.
chiral anomalies one would need the non-traced version. In this paper we will 
determine the non-traced diagonal heat kernel coefficients.
\section{Flat space without gauge connection}
\setcounter{equation}{0}
In a flat space without gauge connection the recursion relations (\ref{ajd}) 
for the heat kernel coefficients become
\bee\label{rec}
(x^\m\pa_\m + j)\, a_j\= j\,\big(\pa^2+X) a_{j-1}
\quad ,\quad [a_0] \= I
\ene
Here we have set $x\pr$ to zero (we will not differentiate at $x\pr$).
We are mostly interested in the diagonal values of the heat kernel 
coefficients. However, it is easy to see that this in turn requires knowledge 
of some derivatives of preceding heat kernel coefficients on the diagonal. 
Indeed, taking the diagonal value of (\ref{rec}) yields
\bee\label{[aj]}
[a_j] \= [\pa^2 a_{j-1}]\,  + X\,[a_{j-1}]
\ene
Thus in particular 
\bee
[a_1] \= X
\ene
but for $j>1$ we must first find $a_{j-1}$ and $\pa^2 a_{j-1}$ on the diagonal.
Applying $\pa^2$ to (\ref{rec}) and then going on the diagonal gives
\bee
[\pa^2 a_{j-1}]\={j-1\ov j+1}\,\Big( [\pa^2\pa^2 a_{j-2}] + X\,[\pa^2 a_{j-2}]
 +2X_,\sp\m\,[\pa_\m a_{j-2}] + X_,\sp\m\sb\m\,[a_{j-2}] \Big)
\ene
Setting $j=2$ and substituting the result in (\ref{[aj]}), we find
\bee
[a_2] \= \sfrac13\,\pa^2 X \,+\, X^2 
\ene
but for $j>2$ we require some new diagonal values of derivatives of $a_{j-2}$.
This recursive procedure ends after $j$ steps since the diagonal value of any 
derivative of $a_0$ vanishes. Differentiating (\ref{rec}) $n$ times and 
taking the diagonal value yields
\bee\label{panaj}
[\pa_{\m_1}\dots\pa_{\m_n} a_j] \=
 -\,{j\ov j+n} [\pa_{\m_1}\dots\pa_{\m_n}\D a_{j-1}]
\ene

To solve these recursion relations in an effective way, we introduce a short 
hand notation for them. Using comma notation for partial derivatives, but 
writing only the {\it number} $n$ of uncontracted derivatives taken, we can 
abbreviate (\ref{panaj}) as
\bee\label{ajn}
\sa_{j,n}\={j\ov j+n}\,\Big(\,\sa_{j-1,(2),n}
                            \,+\,(\X\,\sa_{j-1})_{,n} \,\Big)
\ene
Here, the index $n$ stands for all partial derivatives taken on the left hand 
side of (\ref{panaj}). The {\it sans serif} symbols serve to emphasize that we
are using this short hand notation and at the same time imply evaluation on the
diagonal. The $\pa^2$ on the right hand side has been abbreviated to a 2 in 
parenthesis. Of course one must first distribute the $n$ derivatives over the 
factors of the second term in (\ref{ajn}) before taken the diagonal value.
Doing so yields
\bee\label{ajndis}
\sa_{j,n}\={j\ov j+n}\,\Big(\,\sa_{j-1,(2),n}
                   \,+\,\su_{p=0}^n {n\ch p} \X_{(p}\,\sa_{j-1,n-p)}\,\Big)
\ene
where the parenthesis around the $p$ plus $n-p$ indices imply total 
symmetrization. We write the result of replacing $n$ by $2n$ and contracting 
all derivatives as
\bee
\sa_{j,(2n)}\={j\ov j+2n}\,\Big(\,\sa_{j-1,(2n+2)}\,+\,
\su_{p=0}^{2n} {2n\ch p} \X_{(p}\,\sa_{j-1,2n-p)}\,\Big)
\ene
where the parenthesis now imply not only total symmetrization, but also full 
contraction. Here we introduced the following notation: for any functions 
$F(x)$, $G(x)$ and $H(x)$ we define
\bea\label{distr}
\F_{(2n)} &=& F_{,(\m_1\m_1\dots\m_n\m_n)}(0) \non\\
\F_{(k}\G_{2n-k)} &=& F_{,(\m_1\dots\m_k}(0)\,G_{,\m_{k+1}\dots\m_{2n})}(0)\,
                      \d^{\m_1\m_2}\dots\d^{\m_{2n-1}\m_{2n}} \\
\F_{(k}\G_\el\H_{2n-k-\el)} &=& F_{,(\m_1\dots\m_k}(0)\,
           G_{,\m_{k+1}\dots\m_{k+\el}}(0)\,H_{,\m_{k+\el+1}\dots\m_{2n})}(0)\,
               \d^{\m_1\m_2}\dots\d^{\m_{2n-1}\m_{2n}} \non
\ena
with similar expressions in case the $2n$ derivatives are distributed over yet
more factors. 
Unless otherwise noted, the use of parenthesis as on the left 
hand side of (\ref{distr}) indicates that the $2n$ derivatives are to be 
totally symmetrized {\it and} fully contracted. Note that
\bee\label{F2n}
\F_{(2n)}\= (\pa^2)^n F(0) 
\ene
so in this case the symmetrization in (\ref{distr}) is superfluous (it is
relevant when a gauge connection is present: see (\ref{Xj})). The notation of 
(\ref{F2n}) is due to Avramidi \cite{Avr} (evaluation at the origin is not 
implied there). We generalized it here by allowing distribution of the 
contracted and symmetrized derivatives over any number of factors. 
% TEST: $$\frac23\ \fracm23\ \sfrac23\ \bfrac23\ \afrac23\ {2\ov 3}$$
Note that e.g.
\bea\label{FGex}
\F_{(1}\G_{1)} &=& F_{,\m}(0)\, G_{,\m}(0) \non\\ 
\F_{(1}\G_{3)} &=& F_{,\m}(0)\, G_{,\m\n\n}(0) \= \F_{(1}\G_{1)(2)} \\ 
\F_{(2}\G_{2)} &=& \sfrac13 F_{,\m\m}(0)\, G_{,\n\n}(0) 
                 + \sfrac23 F_{,\m\n}(0)\, G_{,\m\n}(0) \non 
\ena
The general rule for such a reduction is given in (\ref{combi}).
\subsection{Expansion in derivatives}
We first discuss some generic features of the derivative expansion, which can 
be understood without solving the recursion relations. Using only dimensional 
analysis, it follows that through second order in derivatives the diagonal 
values of the heat kernel coefficients must have the structure
\bee\label{dima}
[a_j] = \a_j X^j\ +\su_{k=1}^{j-1}\a_{j\,k}\,X^{k-1} X_{,\m\m} X^{j-k-1} 
+\su_{k=1}^{j-2}\su_{\el=1}^{j-k-1} \a_{j\,k\el}\,
  X^{k-1} X_{,\m} X^{j-k-\el-1} X_{,\m} X^{\el-1} +\dots
\ene
The $\a$'s are numerical coefficients which, in order for (\ref{dima}) to be
hermitian, must satisfy
\bee
\a_{j\ j-k}\= \a_{j\,k}\quad ,\quad\a_{j\,\el\,k}\= \a_{j\,k\,\el}
\ene
These expectations are borne out by our results. In particular, all terms in 
(\ref{dima}) will turn out to have nonvanishing coefficients. At fourth order 
in derivatives, dimensional analysis allows 11 ways of distributing the 
derivatives, namely
\bea\label{fourd}
& X_{,\m\m\n\n} &\non\\
&X_{,\m\m\n} X_{,\n}\quad ,\quad 
 X_{,\m\m} X_{,\n\n} \quad ,\quad X_{,\m\n} X_{,\m\n}& \\
&X_{,\m\m} X_{,\n}X_{,\n}\quad ,\quad X_{,\n} X_{,\m\m} X_{,\n}\quad ,\quad
 X_{,\m\n} X_{,\m}X_{,\n}\quad ,\quad X_{,\m} X_{,\m\n} X_{,\n}& \non\\ 
&X_{,\m} X_{,\m} X_{,\n} X_{,\n}\quad ,\quad 
 X_{,\m} X_{,\n} X_{,\m} X_{,\n}\quad ,\quad
 X_{,\m} X_{,\n} X_{,\n} X_{,\m}& \non
\ena
Here and through (\ref{overlap}) below, we list only equivalence classes of
Lorentz scalars, where two scalars are considered equivalent if they become
equal upon omission of all undifferentiated factors $X$ and/or reversing the 
order of the factors\footnote{By working with real fields only, we can always 
arrange $X$ to be a real symmetric matrix. The same is true for its derivatives
and $\{F\}$ as in (\ref{brac}) then means that, unless $F$ is a palindrome, we 
must add to $F$ its transpose.}. We therefore do not write the curly 
brackets as in (\ref{brac}).

We now claim that two of the Lorentz scalars in (\ref{fourd}) can not appear 
in the heat kernel coefficients. To prove our assertion, we need not solve the
recursion relations explicitly. It suffices to keep track of the way the 
derivatives get distributed over the various factors as the recursion proceeds.
Hence, we omit numerical factors as well as any undifferentiated matrix $X$. 
We may even drop the ordering labels $j$ respectively $j-1$ in (\ref{ajndis}) 
and thus abbreviate $\sa_{j,n}$ to $\sa_n$, where $n$ is the number of 
derivatives. Writing only Lorentz scalars, we thus have the following chain of
substitutions
\bea\label{chain0}
\sa &\ra & \sa_{(2)} \non\\
\sa_{(2n)} &\ra & \sa_{(2n+2)} +\su_{p=1}^{2n}\X_{(p}\sa_{2n-p)} 
                      \quad ,\quad n\geq 1\\
\X_{(p}\sa_{2n-p)} &\ra & \X_{(p}\sa_{2n-p)(2)} 
                    +\su_{q=1}^{2n-p}\X_{(p}\X_q\sa_{2n-p-q)}
                      \quad ,\quad p=1\dots 2n \non
\ena
etc. We truncate this hierarchy at level $2N$ in derivatives, i.e. we drop any
term with more than $2N$ derivatives (note that for the $j$-th coefficient, 
$N\leq j-1$). To be definite, consider the case $N=2$ where the first few 
steps are
\bea\label{ex0}
\sa &\ra & \sa_{(2)} \non\\
    &\ra & \sa_{(4)} +\X_{(1}\sa_{1)} +\X_{(2)}\sa \\
    &\ra &  \X_{(4)} +\X_{(3}\sa_{1)} +\X_{(2}\sa_{2)} +\X_{(1}\sa_{3)}
                     +\X_{(1}\X_{1)}\sa +\X_{(2)}\sa \non
\ena
We used\footnote{Note that in general we have 
$\X_{(p}\sa_{2n-p)(2)}\neq\X_{(p}\sa_{2n-p+2)}$. Thus, before being able to 
iterate again, we must `remove the box'. See (\ref{mobo}) for an example and 
app. B for the general solution. Since we shall not go beyond $N=2$ this 
complication is irrelevant here.} that $\X_{(1}\sa_{1)(2)} = \X_{(1}\sa_{3)}$, 
see (\ref{FGex})
Iterating once more, we see that at fourth order in derivatives a generic heat
kernel coefficient can contain only the following 9 equivalence classes of
Lorentz scalars
\bea\label{nine}
&\X_{(4)}& \non\\
&\X_{(3}\X_{1)}\quad ,\quad \X_{(2}\X_{2)}\quad,\quad\X_{(2)}\,\X_{(2)}&
\non\\
&\X_{(1}\X_2\X_{1)}\quad,\quad \X_{(2}\X_1\X_{1)}\quad ,\quad 
                                          \X_{(2)}\X_{(1}\X_{1)} & \\
&\X_{(1}\X_1\X_1\X_{1)}\quad ,\quad \X_{(1}\X_{1)}\,\X_{(1}\X_{1)}& \non 
\ena
Comparison with (\ref{fourd}) shows that expressions with overlapping sets of 
symmetrized and contracted derivatives, namely
\bea\label{overlap}
&\X_{(1}\X^{(2)}\X_{1)} \equiv X_{,\n} X_{,\m\m} X_{,\n}&    \\
&\X_{(1}\X^{(1}\X^{1)}\X_{1)} \equiv X_{,\m} X_{,\n} X_{,\n} X_{,\m}
\ \quad {\rm or} \quad\ 
\X_{(1}\X^{(1}\X_{1)}\X^{1)} \equiv X_{,\m} X_{,\n} X_{,\m} X_{,\n}& \non
\ena
are absent. This holds for all heat kernel coefficients (recall that we dropped
any undifferentiated matrix $\X$). The absence of the first (second and third)
entry in (\ref{overlap}) can be first observed in $\sa_5$ (respectively 
$\sa_6$). This may hence be relevant to the short distance behavior and 
anomalies of quantum field theories in ten or more dimensions. We conclude 
that at fourth order in derivatives only 9 of the 11 {\it a priori} allowed 
scalars appear. This is only a small reduction in the number of terms, but at 
higher orders the relative number of such absent Lorentz scalars increases 
rapidly. At sixth order in derivatives we find that dimensional analysis and 
hermiticity would permit 85 different Lorentz scalars similar to those in 
(\ref{fourd}). However, continuing (\ref{ex0}), we discover that only 53 of 
these scalars can actually appear in the heat kernel coefficients.

We now give our explicit solution for $\sa_j$ through fourth order in 
derivatives. To have manifest hermiticity, we use here the convention that in 
an $N$-fold sum with $k_1,\dots,k_N$ as summation variables (below 
$k_1,\,k_2,\dots =k,\,\el,\dots$ etc, but skip $o$), the label $k_{N+1}$ has 
by definition the value $(k_N)_{\rm max}-k_N +1$ (so e.g. in the second line
below, $\el\equiv j-k$).
\bea\label{deriv}
&&\sa_j \= \X^j \non\\
&&
+\su_{k=1}^{j-1} {k\el\ov j+1}\,\X^{k-1}\X_{(2)}\X^{\el-1} \non\\
&&
+\su_{k=1}^{j-2}\su_{\el=1}^{j-k-1} {2k\el\ov j+1}\,
\X^{k-1}\X_{(1}\X^{m-1}\X_{1)}\X^{\el-1} \non\\
&&
+\su_{k=1}^{j-2}{k(k+1)\el(\el+1)\ov 2(j+1)(j+2)}\,
\X^{k-1}\X_{(4)}\X^{\el-1} \non\\
&&
+\su_{k=1}^{j-3}\su_{\el=1}^{j-k-2} \Big( {k\el m\ov j+1}\,
 \X^{k-1}\X_{(2)}\X^{m-1}\X_{(2)}\X^{\el-1} \non\\
&&\qquad\qquad
+ {3k(k+1)\el(\el+1)\ov (j+1)(j+2)}\,
   \X^{k-1}\X_{(2}\X^{m-1}\X_{2)}\X^{\el-1} \non\\
&&\qquad\qquad
+ {2k(k+1)\el(\el+m+1)\ov (j+1)(j+2)}\,
 \{\X^{k-1}\X_{(3}\X^{m-1}\X_{1)}\X^{\el-1}\} \Big) \non\\
&&
+\su_{k=1}^{j-4}\su_{\el=1}^{j-k-3}\su_{m=1}^{j-k-\el-2} \Big( 
{2k\el m\ov j+1}\,
 \{\X^{k-1}\X_{(2)}\X^{m-1}\X_{(1}\X^{n-1}\X_{1)}\X^{\el-1}\} \non\\
&&\qquad\qquad
+{6k(k+1)\el(\el+n+1)\ov (j+1)(j+2)}\,
 \{\X^{k-1}\X_{(2} \X^{m-1}\X_1\X^{n-1}\X_{1)}\X^{\el-1}\} \non\\
&&\qquad\qquad
+ {6 k(k+m+1)\el(\el+n+1)\ov (j+1)(j+2)}\,
 \X^{k-1}\X_{(1}\X^{m-1}\X_2\X^{n-1}\X_{1)}\X^{\el-1} \Big)\non\\
&&
+\su_{k=1}^{j-5}\su_{\el=1}^{j-k-4}\su_{m=1}^{j-k-\el-3}
 \su_{n=1}^{j\! -\! k\! -\!\el\! -\! m\! -\! 2}
\Big( {4k\el p\ov j+1}\,
 \X^{k-1}\X_{(1}\X^{m-1}\X_{1)}\X^{p-1}\X_{(1}\X^{n-1}\X_{1)}\X^{\el-1} \non\\
&&\qquad\qquad
+ {12 k(k+m+1)\el(\el+n+1)\ov (j+1)(j+2)}\,
 \X^{k-1}\X_{(1}\X^{m-1}\X_1\X^{p-1}\X_1\X^{n-1}\X_{1)}\X^{\el-1} \Big)\non\\
&&
+\ O(\pa^6) 
\ena
With the restriction of at most four derivatives, this result constitutes an 
explicit solution for {\it all} heat kernel coefficients in flat space without
a gauge connection. It is complete for $j\leq 3$ and yields all terms but one 
in $\sa_4$. To find the coefficient of the `missing' $\X_{(6)}$ term in $\sa_4$
it is best to expand in powers of the matrix potential. This will be presented 
in the next subsection. Also note that the last two terms in (\ref{deriv})
do not appear until $\sa_6$. Taken together, (\ref{deriv}) and (\ref{Xres})
yield the complete answer for the first six heat kernel coefficients. We refer
to sect 6 for the explicit answers for the first five coefficients (to obtain 
the flat space results, replace each $\Z$ in (\ref{a1-a5}) by an $\X$ and omit
all hats and daggers). 
The result for $\sa_6$ is presented in appendix C.
\subsection{Expansion in the potential}
We return to (\ref{ajn}), now paying attention to the order in the matrix 
potential rather than in derivatives. Thus we are looking for an expansion 
that starts as
\bee
[a_j] \= \a\pr_j (\pa^2)^{j-1} X\ +\ O(X^2)
\ene
$\a\pr_j$ being some $j$-dependent numerical factor.
We will show that the recursion relations yield an expression of the form
\bea\label{Xexp}
&&\sa_j = \a\pr_j \X_{(2j-2)}\ 
+\su_k\su_\el \a\pr_{jk\el} \X_{(2k-\el}\X_{\el)(2j-2k-4)} \non\\
&&
\quad + \su_k\su_\el\su_m\su_n\su_p \a\pr_{jk\el mnp}
\X^{\phantom{p}}_{(2k-m}\X_{m-n\phantom{)}}^{(2\el-p}\X^{p)}_{n)(2j-2k-2\el-6)}
\ +\, O(\X^4)
\ena
where we used our shorthand notation (the exponents $2\el -p$ and $p$ on the
second and third factor of the last term also count derivatives).
This is not manifestly hermitian and verifying hermiticity 
thus provides a strong check on our results for the $\a\pr$ coefficients. 
Furthermore, at third order in $\X$, (\ref{Xexp}) shows overlapping sets of 
derivatives. The results of the previous subsection imply that it must
be possible to remove these overlapping terms. Below we show how this can be 
done.

For convenience, we repeat our starting point, eq (\ref{ajn}) 
\bee
\sa_{j,n}\={j\ov j+n}\,\Big(\,\sa_{j-1,n,(2)}
                            \,+\,(\X\,\sa_{j-1})_{,n} \,\Big)
\ene
The first term on the right hand side, which is of zeroth order in $\X$, can be
eliminated by substituting for it from the left hand side, i.e. replace 
$j\ra j-1,\,n\ra n+2$ and contract one pair of derivatives
\bee
\sa_{j-1,n,(2)}\= {j-1\ov j+n+1}\,\Big(\,\sa_{j-2,n,(4)}\,
+\,(\X\,\sa_{j-2})_{,n,(2)} \,\Big)
\ene
Iterating this yields
\bee
\sa_{j,n}\=\su_{k=0}^{j-1}\ (\X\sa_{j-k-1})_{,n,(2k)} \,
\prod_{q=0}^k {j-q\ov j+n+q}\=\su_{k=0}^{j-1}\ {{j\ch k+1}\ov {j+k+n\ch k+1}}
\ (\X\,\sa_{j-k-1})_{,n,(2k)} 
\ene
Clearly, each term is now at least of first order in $\X$. We prefer to reverse
the order of summation and write this as
\bee
\sa_{j,n}\=\su_{k=0}^{j-1} C_{j\,k}^n\,(\X\sa_k)_{,n,(2j-2k-2)} 
\ene
where we defined the combinatorical coefficients $C$ by
\bee\label{C}
C_{j\,k}^n \= {{j\ch k}\ov {2j-k+n-1\ch j-k}}
\ene
Now separate off the $k=0$ term and distribute the derivatives over $\X\sa_k$ 
to obtain
\bee
\sa_{j,n}\= C_{j\,0}^n\,\X_{n,(2j-2)}
+\su_{k=1}^{j-1}\su_{p=0}^{2j'+n} {2j'+n\ch p} C_{j\,k}^n\,
\X_{(2j'+{\hat n}-p} \sa_{k,p)} 
\ene
with $j'\equiv j-k-1$. The set of indices $\m_1\dots\m_n$ is labeled ${\hat n}$
here to indicate that the elements of this set are to be included in the 
indicated symmetrization, but they remain uncontracted. By substituting for 
$\sa_{k,p}$ from the left hand side we find
\bea
&&\sa_{j,n} \= C_{j\,0}^n\,\X_{n,(2j-2)}
      +\su_{k=1}^{j-1}\su_{p=0}^{2j'+n} {2j'+n\ch p}
 C_{j\,k}^n C_{k\,0}^p\,\X_{(2j'+{\hat n}-p}\X_{p)(2k-2)} \non\\
&&\quad
+\su_{k=2}^{j-1}\su_{\el=1}^{k-1}\su_{p=0}^{2j'+n} {2j'+n\ch p}
 C_{j\,k}^n C_{k\,\el}^p\,\X_{(2j'+{\hat n}-p} (\X\sa_\el )_{p)(2k')}
\ena
with $k'\equiv k-\el-1$.
This shows explicitly the terms of second order in $\X$. To proceed to third 
order, we take $n=0$ and distribute the derivatives over $\X\sa_\el$. In 
general
requires two binomial sums, one for each of the sets of derivatives marked $p$ 
and $2k'$, and leads to overlapping sets of derivatives as in (\ref{Xexp}).
To avoid this and to allow us to lump the two sets of derivatives together, we
use that for any functions $F(x)$ and $G(x)$
\bea\label{mobo}
& F_{(1}\,G_{1)(2)} &=  F_{(1}\,G_{3)} \non\\
& F_{(2}\,G_{2)(2)} &= \sfrac56 F_{(2}\,G_{4)} +\sfrac16 F_{(2)}\,G_{(4)}
\ena
etc. Choosing $F=X$, $G=Xa_1$ and taking $x=0$ , the first identity shows that
we can trivially lump the derivatives together in computing $\sa_5$ and the 
second identity shows how to achieve the same for $\sa_6$. In appendix B, we 
show how to do this for arbitrary values of $j$. Restricting for simplicity 
here to $j\leq5$, we obtain as our final result
\bea\label{Xres}
&&\sa_j \= {1\ov {2j-1\ch j}}\,\X_{(2j-2)} \ 
  +\su_{k=1}^{[j/2]}\su_{p=0}^{j-2k} {2j'\ch p}
   C_{j\,k}^0 C_{k\,0}^p\,\{\X_{(2j'-p}\X_{p)(2k-2)}\} \non\\
&&\quad
+\su_{k=2}^{j-1} \su_{\el=1}^{k-1} \su_{p=0}^{2j'}
 \su_{q=0}^{2k'+p} {2j'\ch p} {2k'+p\ch q}
 C_{j\,k}^0 C_{k\,\el}^p C_{\el\,0}^q\,
\X_{(2j'-p}\X_{2k'+p-q}\X_{q)(2\el-2)} \non\\
&&\quad
+ O(\X^4) 
\ena
with $j'\equiv j-k-1$, $k'\equiv k-\el-1$ and the $C$-symbols were defined in 
(\ref{C}). Note that we rewrote the second order terms in manifestly hermitian
form, the range of the double sum having been correspondingly restricted (we 
use $[n]$ to denote the integer part of $n$).
Expression (\ref{Xres}) is one of the main results of this paper.
\section{Flat space with gauge connection}
\setcounter{equation}{0}
We will show that the heat kernel coefficients in the presence of a gauge 
connection can be obtained from their `trivial' counterparts without such a 
connection by making simple covariant substitutions of the kind
\bee
\pa_{\m_1}\dots\pa_{\m_j} X\ \ra\ \de_{(\m_1}\dots\de_{\m_j)} X\, +\, 
F{\rm -dependent\ terms}
\ene
Here partial derivatives are turned into totally symmetrized covariant 
derivatives and in general there are additional field strength dependent terms.

Denote the nonabelian vector connection by $A$. The associated covariant 
derivative and field strength are defined by
\bee
\de_\m \=\pa_\m +A_\m \quad ,\quad F_{\m\n} \= [\de_\m,\de_\n]
\ene
We take $\de$ and thus $F$ to be antihermitian. The heat kernel coefficients 
satisfy
\bee
(x^\m\de_\m + j)\, a_j \= j\,(\de^2 +X)\, a_{j-1}\quad ,\quad [a_0] \= I
\ene
To solve these recursion relations, we find it convenient to work in 
Fock-Schwinger gauge
\bee\label{FSgauge}
x^\m A_\m (x)\=0
\ene
This is equivalent to the requirement that all partial derivatives of the 
gauge connection vanish upon total symmetrization at the origin, i.e. for
$j\geq 1$
\bee\label{symA}
A_{(\m_1\, ,\ \dots\,\m_j)}(0) \= 0 
\ene
In particular, the gauge field vanishes at the origin and for any $n\geq 0$
\bee\label{BOXndA}
(\pa^2)^n \pa\cd A(0) \= 0
\ene
In Fock-Schwinger gauge the recursion relations simplify to ({\it cf} eq 
(\ref{rec}))
\bee
(x^\m\pa_\m + j)\, a_j \= j\,(\pa^2 +\hX)\,a_{j-1}\quad ,\quad [a_0] \= I
\ene
where we defined the differential operator
\bee
\hX \= X + A^\m\sb{,\m} +2A^\m\pa_\m + A^\m A_\m 
\ene
Let $Z$ be its non-differential-operator part, i.e.
\bee\label{Zdef}
Z \= X + A^\m\sb{,\m} + A^\m A_\m
\ene
We will soon need a covariant expression for the partial derivatives of $Z$
at the origin. It is well known and easily verified that in Fock-Schwinger 
gauge the partial derivatives of the gauge field at this point have covariant
values given for $j\geq 1$ by
\bee\label{Aj}
A^\n\sb{,\m_1\dots\m_j}(0) \=
{j\over j+1}\,F_{(\m_1}\sp\n\sb{;\ \dots\,\m_j)}(0)
\ene
Here we use semicolon notation for Yang-Mills covariant derivatives.
Eq (\ref{Aj}) implies
\bea\label{dAjAAj}
A^\n\sb{,\n\m_1\dots\m_j}(0)
&=& {j+1\over j+2}\,F_{(\n}\sp\n\sb{;\m_1\dots\m_j)}(0)\quad ,\quad j\geq 1 \\
A^2\sb{,\m_1\dots\m_j}(0)
&=& \su_{k=1}^{j-1}\,{j\ch k}\,{k(j-k)\ov (k+1)(j-k+1)} 
F_{(\m_1}\sp\n\sb{;\,\dots\m_k}(0) F_{\m_{k+1}}\sp\n\sb{;\,\dots\m_j)}(0)
\ \ \ ,\ \ \ j\geq 2 \non
\ena
If we now define new covariant {\it sans serif\/} symbols $\Y_j$
\bee\label{defY}
\Y_j\sp\n \equiv {j\over j+1}\,F_{(\m_1}\sp\n\sb{;\ \dots\m_j)}(0)
\ene
which are to be treated formally as vectors, then we can abbreviate 
(\ref{Aj}, \ref{dAjAAj}) as
\bea
A^\n\sb{,\m_1\dots\m_j}(0) \=\Y_j\sp\n \quad &,&\quad
A^\n\sb{,\n\m_1\dots\m_j}(0)\=\Y_{(\n}\sp\n\sb{;\, j)}
\quad ,\quad j\geq 1 \non\\
A^2\sb{,\m_1\dots\m_j}(0) &=&\su_{k=1}^{j-1}\,{j\ch k}\,\Y_{(k}\Y_{j-k)}
\ \ \ ,\ \ \ j\geq 2 
\ena
We further note that at the origin, due to (\ref{symA}), we can immediately 
covariantize the partial derivatives of the matrix potential as follows
\bee\label{Xj}
\X_j\,\equiv\,X_{,\m_1\dots\m_j}(0) \= X_{;(\m_1\dots\m_j)}(0)
\ene
Here we used the same symbol $\X_j$ as in sect 3 to mean now a totally
symmetrized $j$-fold covariant derivative of the matrix potential at the 
origin. Thus, the desired covariant expression for {\it any} partial derivative
of $Z$ at the origin is
\bee\label{Zj}
\Z_j\,\equiv\, Z_{,\m_1\dots\m_j}(0) \= 
\X_j +\Y_{(\n}\sp\n\sb{;\, j)}\, +\su_{k=1}^{j-1} {j\ch k}\,\Y_{(k}\Y_{j-k)}
\ene
No implicit contractions occur here except for the scalar product between
$\Y_k$ and $\Y_{j-k}$. Replacing $j$ by $2j$ and contracting all indices yields
\bee\label{Z2j}
\Z_{(2j)}\=\X_{(2j)}\,+\su_{k=1}^{2j-1} {2j\ch k}\,\Y_{(k}\Y_{\el)}
         \=\X_{(2j)}\,+\su_{k=1}^j {2j\ch k}\,\{\Y_{(k}\Y_{\el)}\}
\ene
where $\el=2j-k$ is to be understood. In the second expression we used the 
notation of (\ref{brac}) to obtain a manifestly hermitian result. In sect. 5 
we determine the generalization of (\ref{Zj}, \ref{Z2j}) to a curved space.
In that case we do not find a closed expression which holds for all values 
of $j$.
\subsection{Expansion in derivatives}
Returning to our short hand notation, we have 
\bee\label{ajng}
\sa_{j,n}\={j\ov j+n}\,\Big(\,\sa_{j-1,(2),n}\,+\,(\HX\,\sa_{j-1})_{,n} \,\Big)
\ene
Similar to (\ref{chain0}), we now have the following chain of substitutions
(we again omit numerical factors and ordering labels for the heat kernel 
coefficients, but this time we keep undifferentiated matrices $\X$)
\bea\label{chain1}
\sa         &\ra& \sa_{(2)}\, +\, \X\,\sa \non\\
\sa_{(2n)}  &\ra& \sa_{(2n+2)}\, +\, (\HX\,\sa)_{(2n)}\quad ,\quad n\geq 1
\ena
etc. In the first line we used that $\hX$ at the origin equals $\X$. To remove
$\hX$ from the second line of (\ref{chain1}) as well, we use that
at the origin
\bee\label{lem0}
(\HX\,\sa)_{(2n)} \=(\Z\dg\sa)_{(2n)}\quad ,\quad n\geq 0
\ene
Note that the right hand side no longer contains a differential operator. To 
prove this, we use $\hX = Z+2A^\m\pa_\m$ and $Z\dg = Z - 2A^\m\sb{,\m}$ and 
note that for any function $F$ one has (see appendix D for the proof)
\bee\label{AF}
(\pa^2)^n \pa_\m (A^\m F)(0)\= 0 \quad ,\quad  n\geq 0
\ene
Taking for $F$ a heat kernel coefficient, (\ref{lem0}) follows. Adding one 
more step to the hierarchy (\ref{chain1}), we obtain
\bea\label{chain1n}
\sa        &\ra& \sa_{(2)}\,+\, \X\,\sa \non\\
\sa_{(2n)} &\ra& \sa_{(2n+2)}\, +\su_{p=0}^{2n} \Z_{(p}\do\sa_{2n-p)}
                      \quad ,\quad n\geq 1  \\
\Z_{(p}\do\sa_{2n-p)} &\ra& \Z_{(p}\do\sa_{2n-p)(2)} \,
+\,\Z_{(p}\do (\HX\,\sa)_{2n-p)}  \quad ,\quad p=0\dots 2n \non
\ena
Except for $p=0$ and $p=2n$, (\ref{lem0}) can not be used to eliminate $\hX$ 
from the last term. Instead we proceed as follows (write $r$ for $2n-p$ and 
keep binomial coefficients here)
\bea\label{Zhi}
\Z_{(p}\do(\HX\,\sa)_{r)}
&=&\su_{q=0}^r {r\ch q} \Z_{(p}\do\Big(\Z_q\sa_{r-q)}\,+\,
        2\A^\n\sb{q}\sa_{r-q)\n}\Big) \non\\
&=&\su_{q=0}^r {r\ch q} \Z_{(p}\do\Z_q\sa_{r-q)}\ +\su_{q\pr=0}^{r-1} 
           {r\ch q\pr+1} \Z_{(p}\do 2\A^\n\sb{q\pr+1}\sa_{r-q\pr-1)\n} \non\\
&=&\su_{q=0}^r {r\ch q} \Z_{(p}\do \Big(\Z_q\sa_{r-q)} \,+\,
                    {r-q\ov q+1}\,2\A^\n\sb{q+1}\sa_{r-q-1)\n}\Big) \non\\
&\equiv &\su_{q=0}^r {r\ch q} \Z_{(p}\do \HZ_q\sa_{r-q)} 
\ena
In the second line we shifted the summation index\footnote{\mbox{The 
$q\pr=-1$ term is absent because the gauge connection vanishes at the origin, 
see (\ref{symA})}.} $q$ so as to 
collect terms with the same number of derivatives on the heat kernel 
coefficient. In the third line we can use (\ref{Aj}) to replace $\A_{q+1}$ by 
$\Y_{q+1}$ and thus obtain a covariant answer. The last line defines $\HZ_q$ 
in terms of the previous line. The hatted $\Z$ is designed to absorb the new 
field strength dependent term (this term 
exists even for $q=0$ and we will write $\HZ$ for $\HZ_0$. In case $r=0$ too, 
$\HZ_0=\X$). With this definition of $\HZ$ we can replace the last line of the
hierarchy (\ref{chain1n}) by
\bee
\Z_{(p}\do\sa_{2n-p)}\ \ra\ \Z_{(p}\do\sa_{2n-p)(2)} \,
+\su_{q=0}^{2n-p} \Z_{(p}\do\HZ_q\sa_{2n-p-q)} \quad ,\quad p=0\dots 2n
\ene
Iterating once more does not bring new features\footnote{Other than those 
remarked upon in footnote 6.} and we arrive at the following conclusion.\\
\\
\noindent
{\it The diagonal values of the heat kernel coefficients in the presence of a 
gauge connection are obtained from those without this gauge connection by
making the following covariant substitutions in (\ref{deriv})}
\bea\label{Subs}
\X_{(2j)}              &\ra& \Z_{(2j)} \non\\
\X_{(j}\X_{k)}         &\ra& \Z_{(j}\do\Z_{k)} \non\\
\X_{(j}\X_n\X_{k)}     &\ra& \Z_{(j}\do\HZ_n\Z_{k)} \\
\X_{(j}\X_p\X_n\X_{k)} &\ra& \Z_{(j}\do\HZ_p\HZ_n\Z_{k)} \non
\ena
{\it etc, where $j,k\geq 1$, $n,p,\dots\geq 0$, $\Z_j$ is defined in (\ref{Zj})
and the $\HZ$ act to their right as follows}
\bea\label{Zhatdef}
\HZ_n\Z_{k)} &=&\Z_n\Z_{k)}\, +\,{2\ov n+1}\, k\,\Y_{n+1}\sp\n\Z_{k-1)\n}\non\\
\HZ_p\HZ_n\Z_{k)} &=& \Z_p\HZ_n\Z_{k)}\, +\,{2\ov p+1}\,\Y_{p+1}\sp\n
            (n\HZ_{n-1}\sp\n \Z_{k)} + k\HZ_n\Z_{k-1)\n})
\ena
\\
Since they are unaffected, we omitted $\Z_{(j}\do\dots$ from the left of each 
term in (\ref{Zhatdef}), the dots representing possible further $\HZ$. Note 
that $\HZ_n$ acts on each one of the $k$ indices on $\Z_k$ in the same way, 
hence the factor $k$. Similarly $\HZ_p$ acts on $n$ plus $k$ indices, etc. 
Thus, upon `turning on the gauge field', (\ref{deriv}) becomes
\bee\label{gauged}
\sa_j \= \X^j\,
+\su_{k=1}^{j-1} {k\el\ov j+1}\,\X^{k-1}\Z_{(2)}\X^{\el-1} 
+\su_{k=1}^{j-2}\su_{\el=1}^{j-k-1} {2k\el\ov j+1}\,
\X^{k-1}\Z_{(1}\do\HZ^{m-1}\Z_{1)}\X^{\el-1}
+ \dots
\ene
Note that only an $\X$ which is `sandwiched' between $\X_{(j}$ and $\X_{k)}$ 
gets replaced by a $\HZ$. Formally, the substitutions in (\ref{Subs}) do not 
change the number of terms or their numerical coefficients. In this sense the 
heat kernel coefficients retain their original appearance in the gauging 
process. If we use the curly bracket notation of (\ref{brac}), we should take 
note that for $j\neq k$
\bee
\{\Z_{(j}\do\HZ_n\Z_{k)}\} \= \Z_{(j}\do\HZ_n\Z_{k)} +\Z_{(k}\do\HZ_n\Z_{j)}
\ene
with no dagger on $\HZ_n$ in the second term (this can be shown to be 
hermitian). Finally, repeating the steps of (\ref{ex0}), we find that the 
covariant analogues of (\ref{overlap}) are absent.
\subsection{Expansion in the potential}
Upon replacing $X$ in sect. 3.2 by the differential operator $\hX$ and
retracing our steps, we find that the heat kernel coefficients for the case 
with a gauge connection are obtained from those in (\ref{Xres}) through the 
same substitutions as in (\ref{Subs}). Thus
\bea\label{Zres}
&&\sa_j \= {1\ov {2j-1\ch j}}\,\Z_{(2j-2)} \ 
  +\su_{k=1}^{[j/2]}\su_{p=0}^{j-2k} {2j'\ch p}
   C_{j\,k}^0 C_{k\,0}^p\,\{\Z_{(2j'-p}\do\Z_{p)(2k-2)}\} \non\\
&&\quad
+\su_{k=2}^{j-1} \su_{\el=1}^{k-1} \su_{p=0}^{2j'}
 \su_{q=0}^{2k'+p} {2j'\ch p} {2k'+p\ch q}
 C_{j\,k}^0 C_{k\,\el}^p C_{\el\,0}^q\,
\Z_{(2j'-p}\do\HZ_{2k'+p-q}\Z_{q)(2\el-2)} \non\\
&&\quad
+ O(\Z^4) 
\ena
with $j'\equiv j-k-1$, $k'\equiv k-\el-1$ and the $C$-symbols were defined in 
(\ref{C}). The action of $\HZ_n$ was defined in (\ref{Zhatdef}) (it does not 
act on the set of indices labeled $(2\el-2)$). 
\section{Curved space}
\setcounter{equation}{0}
In this section we shall generalize our results to a curved space with metric 
$g_{\m\n}$. Riemann normal coordinates $x^\m$ can be defined by\footnote{
We use the same symbol for normal coordinates as for general coordinates.
This should not cause confusion. In these coordinates and at the origin there 
is no need to distinguish co- from contravariant indices. We exploit this to
position indices in such a way as to cause minimal clutter. Normal coordinates
$x^\m$ may alternatively be defined in terms of the affine connection by 
$x^\m x^\n \Gamma_{\m\n}^\l (x) = 0$ (which is equivalent to 
$\Gamma_{(\m_1\m_2 ,\,\dots\m_j)}^{\ \l}(0) = 0$ for $j\geq 2$).
Note the similarity with the Fock-Schwinger gauge (\ref{FSgauge}).}
\bee\label{Rnc}
g_{\m\n}(0) \= \d_{\m\n}\quad ,\quad x^\m g_{\m\n}(x) \= x^\m g_{\m\n}(0) 
\ene
which is equivalent to
\bee\label{gsym0}
\quad g_{\m\n}(0) \= \d_{\m\n}\quad ,\quad
g_{\m (\m_1 ,\m_2\dots\m_j)}(0) \= 0 \quad ,\quad j\geq 2
\ene
These properties also hold for the inverse metric $g^{\m\n}$.
Taking the origin of our normal coordinate system to coincide with the point
$x\pr$, we have
\bee\label{signc}
\s(x,0)\= \sfrac12\, x^2 \= \sfrac12\, x^\m x^\n\d_{\m\n}\quad ,\quad
{\cal D}(x,0) \= g(x)^{-1/2}
\ene
and DeWitt's ansatz (\ref{hkc}) for the heat kernel becomes
\bee\label{hkn}
K(x,0;\t) \= (4\p\t)^{-d/2} g(x)^{-1/4}\, e^{-x^2/4\t}
\su_{j=0}^\infty  a_j(x,0) \,{\t^j\ov j!}
\ene
Using (\ref{signc}), the recursion relations (\ref{ajd}) become
\bee\label{curvrec}
(x^\m\pa_\m + j)\, a_j \= j\,(\pa_\m g^{\m\n}\pa_\n + 2A^\m\pa_\m + Z)\,a_{j-1}
\quad ,\quad [a_0]\= I
\ene
Here
\bee
Z\,\equiv\, \ZM + \ZS \quad ,\quad 
\ZM\,\equiv\, X + A^\m\sb{,\m} + A^\m A_\m\quad ,\quad
\ZS\,\equiv\,\sfrac12 B_,\sp\m\sb\m -\sfrac14 B_,\sp\m B_{,\m}
\ene
where we defined
\bee
B \,\equiv\, \ln\,{\cal D} \= -\,\ln\,\sqrt{g}
\ene
Note the position of the explicit inverse metric in (\ref{curvrec}). The 
quantities $\ZM$ and $\ZS$ are the matrix and scalar parts of $Z$ respectively.
Also note the similarity between $\ZM-X$ and $\ZS$. In the flat space limit 
$\ZS$ vanishes and $\ZM$ reduces to the quantity earlier defined as $Z$ in 
(\ref{Zdef}). We now reserve the symbol $Z$ to mean the sum of $\ZM$ and $\ZS$.
\subsection{Scalar part}
We write $\ZS$ with explicit inverse metric as
\bee\label{Zscalar}
\ZS\=\sfrac12\,(g^{\m\n} B_{,\n} )_{,\m}\,-\,\sfrac14\,B_{,\m} g^{\m\n} B_{,\n}
\ene
Our task is to find a covariant expression for the partial derivatives of $\ZS$
at the origin of the normal coordinate system. This in turn requires the
expansion for the (inverse) metric in normal coordinates, which is a 
well-known problem. We find it easier to work with the vielbein, the metric 
being defined as usual by
\bee
g_{\m\n} \= e_\m\sp{a} e_\n\sp{b}\,\d_{ab}
\ene
where $a$ and $b$ are tangent-space indices.
A recursive formula for the vielbein in normal coordinates was given in 
\cite{ABC}. Here we shall give its solution to all orders in the curvature.
Using matrix notation for the vielbein, i.e.
\bee
(\E_j)_\m\sp{a} \,\equiv\, e_\m\sp{a}\sb{,\m_1\dots\m_j}(0)
\ene
and defining the following {\it sans serif} curvature symbols\footnote{Except
for the normalization factor, this agrees with the definition in \cite{Avr}.
See (\ref{Rcon}) for our curvature conventions.}
(compare with (\ref{defY}))
\bee\label{defK}
\K_j\sp{\m\n}\={j-1\over j+1}\,
R_{(\m_1}\sp\m\sb{\m_2}\sp\n\sb{;\dots\m_j)}(0)\quad ,\quad j\geq 2\quad . 
\ene
we can write the recursive formula of \cite{ABC} for the vielbein as follows
\bee\label{recE}
\E_0\=\I\quad,\quad \E_1\= 0 \quad ,\quad
\E_j\= -\,\K_j\, -\,\su_{k=2}^{j-2} {j\ch k}\,{k(k+1)\ov j(j+1)}\,
\K_k\E_\el\ \ , \ \ j\geq 2
\ene
where $\I$ is the $d\times d$ unit matrix and total symmetrization on the 
$k+\el\equiv j$ indices is implied. We treat $\K_j$ as a symmetric matrix in 
the index pair $\m\n$. Taking the trace of such a matrix yields
\bee\label{trKj}
\tr[\K_j] \= {j-1\over j+1}\,
R_{(\m_1\m_2;\dots\m_j)}(0)\quad ,\quad j\geq 2 
\ene
where the Ricci tensor and its covariant derivatives appear. Note that we 
traced over the world indices. This is not to be confused with the trace over 
field or `bundle' indices which does not occur in this paper. In taking the 
trace over a product of $\K$-matrices, say $\tr[\K_j\K_k\K_\el]$, total
symmetrization on the $j+k+\el$ indices is implied. Such traces appear in our 
final expression for the heat kernel coefficients in curved space, e.g.
\bee\quad\tr[\K_{(2}\K_{2)}] \= 
\sfrac19\,R_{(\k}\sp\m\sb\k\sp\n(0)\, R_\l\sp\n\sb{\l)}\sp\m(0)\=
\sfrac1{27} R^{\k\l} R_{\k\l} +\sfrac1{18} R^{\k\l\m\n} R_{\k\l\m\n}
\ene
where the parenthesis on the left hand side imply symmetrization and
contraction (see (\ref{combi}) for the general case). In this way we 
maintain an index-free notation. Returning now to the recursion formula 
(\ref{recE}), the first few cases are
\bee
\E_2 = -\K_2\quad,\quad
\E_3 = -\K_3\quad,\quad
\E_4 = -\K_4 +\sfrac95 \K_{(2}\K_{2)} \quad ,\quad
\E_5 = -\K_5 + 2\K_{(2}\K_{3)} + 4\K_{(3}\K_{2)} 
\ene
where the parenthesis mean symmetrization only.
Note that for $j\geq 5$ the $\E_j$ are not symmetric matrices.
By iteration of (\ref{recE}) we obtain as solution to all orders
\bea\label{solE}
\E_j &=& -\,\K_j\, -\su_{n=1}^{[j/2]-1} (-1)^n
\su_{k_1=2}^{j-2n}\,\su_{k_2=2}^{j_2-2n+2}\!\!...\!
\su_{k_n=2}^{j_n-2} {j\ch k_1,..,k_n} \Big(
 \prod_{i=1}^n {k_i(k_i+1)\ov j_i(j_i+1)} \K_{k_i}\Big) \K_{j_{n+1}}\non\\
j_1\! &\equiv &\! j\quad ,\quad 
j_i\equiv j -\su_{\el=1}^{i-1} k_\el \quad ,\quad i\geq 2
\ena
where total symmetrization of the $j$ indices is to be understood. 
To the best of our knowledge, such an explicit expression for the vielbein in 
normal coordinates was not available in the literature up to now\footnote{We 
have been informed by C. Schubert and U. M\"uller that they have obtained 
results equivalent to (\ref{solE}). Details will be published elsewhere 
\cite{MSV}.}.
For the inverse metric it follows that
\bee\label{invmet}
\sg\sp{\m\n}\sb{j} \= 2\,\K_j\sp{\m\n}
+\,2\su_{k=2}^{[j/2]} {j\ch k}\,
\a_{k\el}\,\{\K_k\K_\el\}\sp{\m\n}\ +\, O(\K^3) \quad ,\quad
\a_{k\el} \,\equiv\, 1\,+\,{k\el\ov j(j+1)}
\ene
where $\el\equiv j-k$ (see (\ref{alphaord}) for our summation conventions). 
Finding the heat kernel coefficients through fourth order in $\K$ turns out 
to require knowledge of the inverse metric through second order only. See 
appendix E.
Explicit expansions for the metric and its inverse to some finite order in
normal coordinates are well known, see e.g. \cite{Sak,McL} and our eq 
(\ref{invmetexpl}). We next use
\bee
B(x) =\, -\tr\ln E\= -\su_{k=1}^\infty {1\ov k}\tr[(I-E)^k]
\ene
Thus $B$ and its first derivative vanish at the origin and for $j\geq 2$, after
collecting terms of the same degree in $\K$, we find ({\it cf} (\ref{Bexpl}))
\bea\label{Bj}
\B_j &=& \tr[\K_j]\,
+\su_{k=2}^{[j/2]} {j\ch k} \b_{k\el} \tt[\K_k\K_\el]\ 
+\su_{k=2}^{[j/3]}\su_{\el=k}^{[(j-k)/2]} {j\ch k,\el}
 \b_{k\el m}\tt[\K_k\K_\el\K_m] \non\\
&&\quad
+\su_{k=2}^{[j/4]}\su_{\el=k\pr\pr}^{[j/2]-k}\su_{m=k}^{j-k-2\el} 
{j\ch k,\el,m} \b_{k\el mn}\tt[\K_k\K_\el\K_m\K_n]\ 
+\ O(\K^5) 
\ena
where $\tt$ is a weighted trace defined below and the lower limit $k\pr\pr$ 
means: when $\el=k$, reduce the upper limit for the sum over $m$ to $[j/2]-k$.
The coefficients are given by
\bea\label{betas}
\b_{k\el}   &=& 1\,-\,P_2\Big[{k(k+1)\ov j(j+1)}\Big] \={2k\el\ov j(j+1)}\non\\
\b_{k\el m} &=& 2\,-\,P_3\Big[{k(k+1)(j+\el+1)\ov j(j+1)(j-\el+1)}\Big] \=
{4k\el m (j^2+4j+3+k\el+\el m+mk)\ov j(j+1) D_{k\el} D_{\el m} D_{mk}} \non\\
\b_{k\el mn}&=& 2\,-\,P_4\Big[{k(k+1)\ov (k+\el) D_{k\el}}
\Big(1-{n(n+1)(j+m+1)\ov j(j+1)D_{k\el n}}-{m(m+1)\ov 2(m+n)D_{mn}}\Big)\Big] 
\non\\
&&\quad D_{k\el} \equiv k+\el+1 \quad ,\quad D_{k\el m} \equiv k+\el+m+1
\ena
Here $P_N$ denotes the group consisting of all cyclic and anticyclic 
permutations on $N$ objects (so $\abs{P_2} = 2,\,\abs{P_N} =2N$ for 
$N>2$) and its action is defined by
\bee
P_N[f(k_1,\dots,k_N)]\=\su_\p  f(k_{\p(1)},\dots,k_{\p(N)})
\quad ,\quad j\equiv\su_{i=1}^N k_i
\ene
where the sum runs over all cyclic and anticyclic permutations. We used here 
that the trace of a product of symmetric matrices is invariant under such 
permutations in order to simplify the sums. Thus, in $\tr[\K_k\K_\el\K_m]$ the
order of the matrices is irrelevant and we can assume without loss of 
generality that $k\leq\el\leq m$. Similarly, in $\tr[\K_k\K_\el\K_m\K_n]$ we 
can assume $k$ to be the smallest label and $\el\leq n$ (when $\el=k$ we 
arrange $m\leq n$). Ordering subtleties do not occur before fourth order, 
where we must distinguish\footnote{Thus in obtaining the logarithm of the Van 
Vleck-Morette determinant through ninth order in the normal coordinates, see 
(\ref{Bexpl}), we may treat the $\K_j$ as commuting objects !}
between $\tr[\K_2\K_2\K_3\K_3]$ and $\tr[\K_2\K_3\K_2\K_3]$. 
The weighted trace $\tt$ is defined by
\bee\label{symmfac}
\tt[\K_{k_1}\K_{k_2}\dots\K_{k_N}]\= 
    {S\ov \abs{P_N}}\,\tr[\K_{k_1}\K_{k_2}\dots\K_{k_N}]
\ene
with $S$ the total number of distinguishable cyclic and anticyclic
permutations (including the identity) of the product 
$\K_{k_1}\K_{k_2}\dots\K_{k_N}$. We refer to appendix F for a table
containing all symmetry factors for $N\leq 4$.
%, so e.g. for $N=2$
%\bee
%\tt[\K_k\K_k]  \= \sfrac12 \tr[\K_k\K_k] \quad ,\quad
%\tt[\K_k\K_\el]\= \tr[\K_k\K_\el]\ \ \, {\rm for}\,\ \ k\neq\el
%\ene
Substituting (\ref{Bj}) and (\ref{invmet}) into (\ref{Zscalar}) we find
for any $j\geq 0$ but with a cutoff at third order in $\K$
(see appendix E for the details and (\ref{Zs1-5}) for examples)
\bea\label{ZSj}
\Zs_j &=& \sfrac12\tr[\K_{j\,\m\m}]\ 
+\su_{k=2}^j {j+1\ch k} \K_{(k}\sp{\m\n}\tr[\K_{j-k\,\m)\n}]  \non\\
&& 
+\su_{k=1}^{[j/2]\pr} {j\ch k}
\Big( {j+1\ov j+3}\tr[\K_{\m k} \K_{\m\el}]
\, -\,\sfrac12\tr[\K_{\m(k}] \tr[\K_{\el)\m}] \Big) \non\\
&&
+\su_{k=2}^{[{j+2\ov 3}]}\su_{\el=k}^{[{j-k+2\ov 2}]} {j+2\ch k,\el}\,
\sfrac12\b_{k\el m}\tt[\K_{\m k-1}\K_m\K_{\m\,\el-1}] \non\\
&&
-\su_{k=1}^{j-3}\su_{\el=2}^{[{j-k+1\ov 2}]} {j\ch k,\el} {\el\ov j-k+2}
\tr[\K_{\m(k}] \tt[\K_\el \K_{m)\m}] \non\\
&&
-\su_{k=1}^{[j/2]-1}\su_{\el=k\pr}^{j-k-2} {j\ch k,\el} 
\tr[\K_{\m(k}]\,\K_m\sp{\m\n}\tr[\K_{\el)\n}]  \non\\
&&
+\su_{k=4}^j {j+1\ch k} \su_{\el=2}^{[k/2]} {k\ch\el}
\Big( 1 + {\el (k-\el)\ov k(k+1)}\Big) \{ \K_{(\el}\K_{k-\el}\}^{\m\n}
\tr[\K_{j-k\,\m)\n}] \non\\
&&
+\su_{k=2}^{j-2} \su_{\el=2}^{[(j-k+2)/2]} {j+1\ch k,\el}
{2\el\ov j-k+3} \K_{(k}\sp{\m\n}\tt[\K_\el\K_{j-k-\el\,\m)\n}] \non\\
&&
+\ O(\K^4) 
\ena
This expression is complete for $j\leq 5$. The coefficient $\b_{k\el m}$ which
appears here is obtained by replacing $j$ by $j+2$ in (\ref{betas}) and 
$\{\K_k\K_\el\}$ is defined as in (\ref{brac}). The prime on $[j/2]$ in the 
second sum implies division by 2 when $j$ is even and this limit is reached. 
Similarly, the lower limit $\el=k\pr$ means division by 2 when $\el=k$. 
We recall that total symmetrization on {\it all} indices within a given trace 
(except for those being traced over) is to be understood and thus one has e.g.
\bee
\tr[\K_{\m(1}] \tr[\K_3\K_{1)\m}] \= \tr[\K_{\m(1}] \tr[\K_2\K_{2)\m}] 
\ene
Replacing $j$ by $2j$ in (\ref{ZSj}) and contracting fully, the second and last
two terms vanish. Keeping also terms of fourth order in $\K$ we find ({\it cf}
(\ref{Zs(2)-(8)}))
\bea\label{Zs2j}
\Zs_{(2j)} &=& \sfrac12\tr[\K_{(2j+2)}] \non\\
&&
+\su_{k=1}^{j\pr} {2j\ch k}\,\Big( {2j+1\ov 2j+3}\tr[\K_{(k+1}\K_{\el+1)}]
                   \, -\,\sfrac12\tr[\K_{\m(k}] \tr[\K_{\el)\m}] \Big)\non\\
&&
+\su_{k=2}^{[{2j+2\ov 3}]}\su_{\el=k}^{[{2j-k+2\ov 2}]} {2j+2\ch k,\el}\,
\sfrac12\,\b_{k\el m}\tt[\K_{(k}\K_\el\K_{m)}] \non\\
&&
-\su_{k=1}^{2j-3}\su_{\el=1}^{[{2j-k-1\ov 2}]} {2j\ch k,\el}
\g_{\el m} \tr[\K_{\m(k}]\tt[\K_m\K_{\el)\m}]  \non\\
&&
-\,\su_{k=1}^{j-1}\,\su_{\el=k\pr}^{2j-k-2}\, {2j\ch k,\el} 
\tr[\K_{\m(k}]\,\K_m\sp{\m\n}\tr[\K_{\el)\n}] \non\\
&&
+\su_{k=2}^{[{j+1\ov 2}]} \su_{\el=k\pr\pr}^{j-k+1} \su_{m=k}^{2j-k-2\el+2} 
 {2j+2\ch k,\el,m}\,\sfrac12\,\b_{k\el mn} \tt[\K_{(k}\K_\el\K_m\K_{n)}]\non\\
&&
-\su_{k=1}^{2j-5} \su_{\el=1}^{[{2j-k-2\ov 3}]} 
 \su_{m=\el+1}^{[{2j-k-\el\ov 2}]} {2j\ch k,\el,m} \g_{\el mn}
\tr[\K_{\m(k}] \tt[\K_n\K_m\K_{\el)\m}] \non\\
&&
-\su_{k=1}^{[{j-1\ov 2}]} \su_{\el=k+1}^{(j-k)\pr}
 \su_{m=1}^{[{2j-k-\el-1\ov 2}]}  {2j\ch k,\el,m} 2\g_{k\el}\g_{mn}
\tt[\K_{\m(k}\K_\el] \tt[\K_n\K_{m)\m}] \non\\
&&
-\su_{k=1}^{j-2}\,\su_{\el=k\pr}^{2j-k-4} \su_{m=2}^{[{2j-k-\el\ov 2}]}\ \,
{2j\ch k,\el,m}\a_{mn}\tr[\K_{\m(k}]\,\{\K_m\K_n\}^{\m\n}\tr[\K_{\el)\n}]\non\\
&&
-\su_{k=1}^{2j-5} \su_{\el=1}^{[{2j-k-3\ov 2}]} \su_{m=\el+1}^{2j-k-\el-2}
 {2j\ch k,\el,m} 2\g_{\el m}
\tr[\K_{\m(k}]\,\K_n\sp{\m\n} \tt[\K_m\K_{\el)\n}] \non\\
&&
+ \ O(\K^5)
\ena
This expression is complete for $j\leq 3$. The $\a_{mn}$ are defined in 
(\ref{invmet}) (with $j\ra m+n\equiv 2j-k-\el$) and the $\b$-coefficients are 
those of (\ref{betas}) with $j\ra 2j+2$. The $\g$-coefficients are defined by
\bee
\g_{k\el} \,\equiv\,  {\el\ov k+\el+2} \quad ,\quad
\g_{k\el m}\,\equiv\, {k+\el+m+1\ov 2(k+1)}\,\b_{k+1\,\el\, m}
\ene
The upper limit $j\pr$ means: divide by 2 when $k=j$. A lower limit $k\pr$ 
means: divide by 2 when $\el=k$. The upper limit $(j-k)\pr$ means: divide by 2
when $\el=j-k$ and $m=k$. Finally, the lower limit $k\pr\pr$ means: when
$\el=k$, reduce the upper limit for the sum over $m$ to $j-k+1$.
\subsection{Matrix part}
Abbreviate the $j$-th partial derivative of the gauge connection at the origin
by
\bee
\A_j \= A_{\m,\m_1\dots\m_j}(0)
\ene
Then the recursion for the gauge connection is given by (until further notice, 
parenthesis imply symmetrization only)
\bee
\A_0 \= 0\quad ,\quad 
\A_j \=\Y_j +\su_{k=2}^{j-1} {j\ch k} \,{\el+1\ov j+1}\,\E_{(k}\Y_{\el)}
\ \ ,\ \ j\geq 1
\ene
We iterate this and use (\ref{solE}) to obtain as solution to all orders
\bea\label{solA}
\A_j\!\! &=&\!\!\Y_j +\su_{n=1}^{[{j-1\ov 2}]} (-1)^n \su_{k=1}^{j-2n}
\su_{k_1=2}^{j_1-2n+2}\!...\!\!
\su_{k_{n-1}=2}^{j_{n-1}-2} {j\ch k,..,k_{n-1}} {k+1\ov j+1} \Y_k 
\Big(\prod_{i=1}^{n-1} {k_i(k_i+1)\ov j_i(j_i+1)} \K_{k_i}\Big)\K_{j_n} \non\\
j_i &\equiv & j-\su_{\el=0}^{i-1} k_\el\quad ,\quad k_0\,\equiv\, k 
\ena
where symmetrization on the $j$ indices is to be understood. We used that, 
since the $\K_j$ are symmetric matrices, we may write them to the right of the
vector $\Y_{k_0}$ in the reverse order. Through second order in $\K$, 
(\ref{solA}) reads ({\it cf} (\ref{Aexpl}))
\bea\label{expA}
\A_j &=&\Y_j\ -\su_{k=1}^{j-2}{j\ch k} {k+1\ov j+1}\,\Y_{(k}\K_{\el)} \\
&&\quad
+\su_{k=1}^{j-4}\su_{\el=2}^{j-k-2} {j\ch k,\el} 
{(k+1)\el(\el+1)\ov (j+1)(j-k)(j-k+1)}\,\Y_{(k}\K_\el\K_{m)}\ +\,O(\K^3)\non
\ena
We thus find (this is exact for $j\leq 5$; see (\ref{Zm1-5}) for examples)
\bea\label{Zmj}
&&
\Zm_j = \X_j\, +\Y_{(\n}\sp\n\sb{;j)}\, +\su_{k=1}^{j-1} \Y_{(k}\Y_{\el)}\ 
+\su_{k=2}^j{j+1\ch k}{j+k+2\ov j+2}\,\K_{(k}\sp{\m\n}\Y_\m\sp\n\sb{;j-k)}
\non\\
&&
+ \su_{k=1}^{[j/2]-1}\su_{\el=k}^{j-k-2} {j\ch k,\el}\,
\big({m\ov k+m+1} + {m\ov \el+m+1}\big)\, \{\Y_{(k}\K_m\Y_{\el)}\}\non \\
&&
+\su_{k=2}^{j-2}\su_{\el=2}^{j-k} {j+1\ch k,\el}\,\Big({3\el+m+1\ov\el+m+1}
\,-\,{\el(\el+1)\ov (j+2)(k+\el+1)}\,-\,{k(k+1)\ov (k+\el)(k+\el+1)}\Big)\non\\
&&\qquad\qquad\qquad\qquad 
\times(\K_{(k}\K_\el)\sp{\m\n} \Y_\m\sp\n\sb{;j-k-\el)}\ +\ O(\K^3\Y,\K^2\Y^2)
\ena
If we now replace $j$ by $2j$ in (\ref{Zmj}) and contract all indices, the
second, fourth and last term vanish. Including terms of fourth order in
curvatures, we find ({\it cf} (\ref{Zm(2)-(8)}))
\bea\label{Zm2j}
&&
\Zm_{(2j)} =\X_{(2j)}\,+\,\su_{k=1}^j {2j\ch k}\{\Y_{(k}\Y_{\el)}\}\non\\
&&
+ \su_{k=1}^{j-1}\su_{\el=k}^{2j-k-2} {2j\ch k,\el}
\big({m\ov k+m+1} + {m\ov \el+m+1}\big) \{\Y_{(k}\K_m\Y_{\el)}\} \\
&&
+\su_{k=1}^{j-2}\su_{\el=k}^{2j-k-4}\su_{m=2}^{2j-k-\el-2}{2j\ch k,\el,m}
{mn\,N_{k\el m}\ov D_{km} D_{\el n} D_{mn} D_{kmn} D_{\el mn}}
\{\Y_{(k}\K_m\K_n\Y_{\el)}\}\ +\,O(\K^3) \non
\ena
where the $D$-symbols were defined in (\ref{betas}) and 
\bea
N_{k\el m} &=& 5 + 5k\el + 12j + 16jmn\ +\{6m +(k+m)(9m+10n) \non\\
           & &\qquad\quad +k^2(\el+m+2n+2) +(3k+4\el+2m)m^2 +4k\el m\}_S \non\\
&&\{f(k,\el,m,n)\}_S \equiv  f(k,\el,m,n) + f(\el,k,n,m) 
\ena
We recognize the first line of (\ref{Zm2j}) as the flat spacetime result 
(\ref{Z2j}). As we already mentioned there, we are not able to give a closed 
expression to all orders in $\K$ here. However, the terms of $N$-th order in 
$\K$ in $\Zm_{(2j)}$ will be of the form 
\bee
\{\Y_{(k}\K_{m_1}\dots\K_{m_N}\!\Y_{\el)}\}\quad ,\quad
2j \= k+\el +\su_{i=1}^N m_i\quad ,\quad k\leq\el
\ene
and this shows that $\ZM$ has a much simpler structure than $\ZS$.
\subsection{Expansion in derivatives}
We retrace the steps of sect 4.1, starting with (compare with (\ref{ajng}))
\bee\label{ajnc}
\sa_{j,n}\={j\ov j+n}\,\Big(\,\sa_{j-1,}\sp\m\sb{\m,n}\,
           +\,(\HX\,\sa_{j-1})_{,n} \,\Big)
\ene
where now
\bee
\hX \= Z + 2 A^\m\pa_\m \= \ZM +\ZS+ 2 A^\m\pa_\m
\ene
Except for $n=0$ or 1, we can {\it not} replace the explicit pair of indices
in (\ref{ajnc}) by our shorthand notation as in (\ref{ajng}). However,
for even values of $n$ and after full contraction we can use the following 
lemma: for any scalar $F(x)$
\bee\label{lem2}
\F\sp\m\sb{\m\,(2n)} \= \F_{(2n+2)} \quad ,\quad n\geq 0
\ene
This follows immediately from the defining properties of the inverse metric in
normal coordinates, see (\ref{gsym0}). Using this lemma, we obtain the 
following chain of substitutions (omit numerical factors and ordering labels, 
but keep undifferentiated $\Z$'s)
\bea\label{chain2}
\sa        &\ra& \sa_{(2)}\, +\, \Z\,\sa\non\\
\sa_{(2n)} &\ra& \sa_{(2n+2)}\, +\, (\Z\dg\sa)_{(2n)}\quad ,\quad n\geq 1\\
\Z_{(p}\do\sa_{2n-p)} &\ra& \Z_{(p}\do\sa^\m\sb{2n-p)\m} \,
+\,\Z_{(p}\do (\HX\,\sa)_{2n-p)}  \quad ,\quad p=0\dots 2n \non
\ena
etc, where
\bee
\Z\=\Zm +\Zs \=\X +\sfrac12\tr[\K_{(2)}] \=X(0)+\sfrac16 R(0)
\ene
In the second step we used that (\ref{AF}), with the understanding that 
$\pa^2= \pa_\m\pa_\m$, remains true in normal coordinates. We can remove $\hX$
from the last line exactly as in the first three lines of (\ref{Zhi}). However,
the first term on the right hand side in the last line of (\ref{chain2}) yields
a new term as compared to flat space ($r\equiv 2n-p$)
\bea\label{rewr}
\Z_{(p}\do\sa^\m\sb{r)\m} \,-\,\Z_{(p}\do\sa_{r)(2)} &\equiv& 
              \Z_{(p}\do({\sf h}^{\m\n}\sa_{,\vert\n\vert})_{r)\m} \non\\
&=&
-\,{p\ov r+1}\,\Z_{\m(p-1}\do({\sf h}^{\m\n}\sa_{,\vert\n\vert})_{r+1)} \non\\
&=&
-\,{p\ov r+1}\su_{q=0}^{r+1} {r+1\ch q}
         \Z_{\m(p-1}\do {\sf h}^{\m\n}\sb{q}\sa_{r-q+1)\n}  \\
&=&
-\su_{q=0}^r {r\ch q} {p(r-q)\ov (q+1)(q+2)}\,
         \Z_{\m(p-1}\do \sg^{\m\n}\sb{q+2}\sa_{r-q-1)\n} \non
\ena
The first step merely defines ${\sf h}^{\m\n}\equiv\sg^{\m\n}-\d^{\m\n}$.
The vertical bars indicate that the index $\n$ is not to be included in the
indicated symmetrization. In the second step we use that, due to (\ref{gsym0})
\bee
\Z_{(p}\do({\sf h}^{\m\n}\sa_{,\vert\n\vert})_{r\,\m)}\= 0 
\ene
Writing this out with respect to the position of the index $\m$ shows the
equality of the first two lines of (\ref{rewr}). Next we use the binomial 
theorem and in the last step we shift $q$ and use that $h$ and its first 
derivative vanish at the origin. We absorb this new term through a redefinition
of $\HZ_q$ in (\ref{Zhi}) and arrive at the following conclusion.\\
\\
\noindent
{\it The diagonal values of the heat kernel coefficients in curved space are 
obtained from those in flat space by formally making the same substitutions as
in (\ref{Subs}), where $\Z_j$ now stands for $\Zm_j+\Zs_j$ (see (\ref{ZSj}, 
\ref{Zmj}) and the action of $\HZ$ is defined by}
\bee\label{TheoII}
\Z_{(j}\do\HZ_n\Z_{k)}\= \Z_{(j}\do\Z_n\Z_{k)}\, 
                          +\,{2k\ov n+1}\,\Z_{(j}\do\A^\n\sb{n+1}\Z_{k-1)\n}
\, -\,{jk\ov(n+1)(n+2)}\,\Z_{\m(j-1}\do\sg^{\m\n}\sb{n+2}\Z_{k-1)\n}
\ene
{\it etc, where the (contravariant) gauge connection and inverse metric 
follow from (\ref{invmet}) and (\ref{expA}).}\\
\\
\noindent
These redefinitions being understood, the heat kernel coefficients formally 
remain unchanged upon going to curved spacetime.
\subsection{Locally symmetric space}
The results of the previous sections nearly suffice to find $\sa_5$. The few 
missing terms are of fifth order in the Yang-Mills and Riemann curvature.
They are most easily found by considering a locally symmetric space, i.e. a
space with covariantly constant curvatures. In that case one can find closed 
expressions which hold through all orders in these curvatures. This situation 
has been considered in detail by Avramidi \cite{ls}. Here, we merely want to
find those terms in a given heat kernel coefficient for a {\it general} 
curved space which do not contain explicit covariant derivatives. We therefore
do not take account of the consequences of the requirement that the Riemann 
curvature is covariantly constant (i.e. $\de R=0$ would imply $[\de,\de]R=0$).

Thus, consider the case where only $\X_0$, $\Y_1$ and $\K_2$ are 
nonvanishing. Define
\bee
Y_\n(x) \= {1\ov 2} \, F_{\m\n}(0)\,x^\m\quad ,\quad
K_{\m\n}(x) \= {1\ov 3} \,R_{\m\r\n\s}(0)\, x^\r x^\s
\ene
The vielbein is then given to all orders in $K$ by (see (\ref{solE}) 
\bee
E[K] \=\su_{k=0}^\infty {1\ov (2k+1)!} (-3K)^k \= {\sinh\,S\ov S}\quad ,\quad
S\,\equiv\,\sqrt{-3 K}
\ene
Note that the vielbein is an even function of $S$ and hence it depends only on
$K$. Since $K$ is a symmetric matrix, so is $E$. It follows that the inverse 
metric simply equals $E^{-2}$. Differentiating $B= -\tr\ln E$ we find
\bee\label{dB}
B_{,\m}\= {1\ov 2} \tr[ K_{,\m} L] \quad ,\quad
L[K] \equiv {3\ov S}\,\big(\coth S \,-\,{1\ov S}\,\big) 
\ene
where $L$ is an even function\footnote{Except for the extra factor $3/S$,
this is the well known Langevin function.} of $S$. In general, the commutator 
of $K$ and $K_{,\m}$ does not vanish, but the trace insures that (\ref{dB})
holds. Differentiating $L$ we find
\bee
L_{,\m} \= \langle K_{,\m} L\pr \rangle \quad ,\quad
\langle K_{,\m}K^n\rangle \equiv {1\over n}\su_{k=0}^n K^k K_{,\m} K^{n-k}
\ene
With this notation, the result for $\ZS$ in a locally symmetric space reads
\bee\label{ZSloc}
\ZS[K] = \sfrac14 (E^{-2})^{\m\n} \Big(\tr[ K_{,\m\n} L + K_{,\m} L_{,\n} ]
-\sfrac14 \tr[K_{,\m} L]\tr[K_{,\n} L] \Big) 
+\sfrac14 \langle E^{-2} K_{,\m} L\rangle^{\m\n} \tr[ K_{,\n} L ]
\ene
To obtain a similar result for $\ZM$ we note that the covariant 
gauge connection is given by
\bee
A[Y,K] \=\Big({\sinh\,S/2\ov S/2}\Big)^2\,Y
\ene
The contravariant gauge connection is therefore
\bee
A[Y,K] \= ({\rm sech}\,S/2)^2\,Y
\ene
and we obtain the following result for $\ZM$ in a locally symmetric space
\bea\label{ZMloc}
&&\ZM[X,Y,K]\= X-\pa\cd(\tanh\,S/2)^2\,Y +Y\Big({\tanh\,S/2\ov S/2}\Big)^2 Y \\
&&\= X +\pa\cdot (\sfrac34 K +\sfrac38 K^2+\,\dots )Y
 + Y (I +\sfrac12 K +\sfrac{17}{80} K^2 +\sfrac{93}{1120} K^3+\,\dots)Y\non
\ena
\section{Explicit results for $\sa_1$ through $\sa_5$}
\setcounter{equation}{0}
As an application, we give here the explicit results for the diagonal values 
of the first five heat kernel coefficients, obtained from (\ref{Subs}), 
(\ref{Zres}) and (\ref{TheoII}). The number of terms equals 1,2,4,10 and 26 
respectively. The sixth coefficient in flat space can be found in appendix C.
\bea\label{a1-a5}
\sa_1 &=& \Z \non\\
\sa_2 &=& \Z^2 
+\sfrac13\Z_{(2)}  \non\\
\sa_3 &=& \Z^3 
+\sfrac12\{\Z\Z_{(2)}\} 
+\sfrac12\Z_{(1}\do\Z_{1)}
+\sfrac1{10}\Z_{(4)} \non\\
\sa_4 &=& \Z^4 
+\sfrac35\{\Z^2\Z_{(2)}\} 
+\sfrac45\Z\Z_{(2)}\Z
+\sfrac25\Z_{(1}\do\HZ\Z_{1)} 
+\sfrac45\{\Z\Z_{(1}\do\Z_{1)}\} \non\\
&&
+\sfrac15\{\Z\Z_{(4)}\}
+\sfrac25\{\Z_{(1}\do\Z_{3)}\}
+\sfrac25\Z_{(2}\do\Z_{2)} 
+\sfrac15\Z_{(2)}\sp{2} 
+\sfrac1{35}\Z_{(6)} \non\\
\sa_5 &=& \Z^5 
+\sfrac23\{\Z^3\Z_{(2)}\}
+\{\Z^2\Z_{(2)}\Z\}
+\sfrac13\Z_{(1}\do\HZ^2\Z_{1)}
+\sfrac23\{\Z\Z_{(1}\do\HZ\Z_{1)}\}
+\{\Z^2\Z_{(1}\do\Z_{1)}\}\non\\
&&
+\sfrac43\Z\Z_{(1}\do\Z_{1)}\Z 
+\sfrac27\{\Z^2\Z_{(4)}\}
+\sfrac37\Z\Z_{(4)}\Z 
+\sfrac67\{\Z\Z_{(3}\do\Z_{1)}\}
+\sfrac8{21}\{\Z_{(1}\do\HZ\Z_{3)}\} \non\\
&&
+\sfrac{16}{21}\{\Z\Z_{(1}\do\Z_{3)}\}
+\sfrac27  \Z_{(2}\do\HZ\Z_{2)}
+\sfrac67\{\Z\Z_{(2}\do\Z_{2)}\}
+\sfrac13\Z_{(2)}\Z\Z_{(2)}
+\sfrac13\{\Z\Z_{(2)}\sp{2}\} \non\\
&&
+\sfrac97\Z_{(1}\do\HZ_2\Z_{1)}
+\sfrac67\{\Z_{(1}\do\HZ_1\Z_{2)}\}
+\sfrac13\{\Z_{(1}\do\Z_{1)}\Z_{(2)}\}
+\sfrac1{14}\{\Z\Z_{(6)}\}
+\sfrac3{14}\{\Z_{(1}\do\Z_{5)}\}\non\\
&&
+\sfrac5{14}\{\Z_{(2}\do\Z_{4)}\}
+\sfrac5{14}  \Z_{(3}\do\Z_{3)} 
+\sfrac2{21}\{\Z_{(2)}\Z_{(4)}\}
+\sfrac4{21} \Z_{(3)}\do\Z_{(3)}
+\sfrac1{126}\Z_{(8)}
\ena
The $\Z$'s appearing here were defined in the previous sections, but to be 
quite explicit and to avoid misunderstanding of our conventions, we give them 
below. We have $\Z_j\=\Zm_j + \Zs_j$ with the matrix quantities, see 
(\ref{Zmj}), given by
\bea\label{Zm1-5}
\Zm_1 &=&\X_1 +\Y_{(\m}\sp\m\sb{;1)} \non\\
\Zm_2 &=&\X_2 +\Y_{(\m}\sp\m\sb{;2)} +2\Y_{(1}\Y_{1)}
              +\sfrac92 \K_{(2}\sp{\m\n}\Y_{\m)\n} \non\\
\Zm_3 &=&\X_3 +\Y_{(\m}\sp\m\sb{;3)} +3\{\Y_{(1}\Y_{2)}\} 
+\sfrac25\Big( 21 \K_{(2}\sp{\m\n}\Y_\m\sp\n\sb{;1)}
              +16 \K_{(3}\sp{\m\n}\Y_{\m)\n} \Big)\non\\
\Zm_4 &=&\X_4 +\Y_{(\m}\sp\m\sb{;4)} +4\{\Y_{(1}\Y_{3)}\} 
                           +6\Y_{(2}\Y_{2)}+12\Y_{(1}\K_2\Y_{1)} \non\\
&&\quad
+\sfrac53\Big( 8 \K_{(2}\sp{\m\n}\Y_\m\sp\n\sb{;2)}
             + 9 \K_{(3}\sp{\m\n}\Y_\m\sp\n\sb{;1)} 
            + (5 \K_{(4} +9\K_{(2}\K_2)\sp{\m\n}\Y_{\m)\n} \Big) \non\\
\Zm_5 &=& \X_5 +\Y_{(\m}\sp\m\sb{;5)} +5\{\Y_{(1}\Y_{4)}\}
   +10\{\Y_{(2}\Y_{3)}\} + 24\Y_{(1}\K_3\Y_{1)} +9\{\Y_{(1}\K_2\Y_{2)}\} \non\\
&&\quad
+\sfrac37\Big( 45 \K_{(2}\sp{\m\n}\Y_\m\sp\n\sb{;3)}
             +\sfrac{200}3 \K_{(3}\sp{\m\n}\Y_\m\sp\n\sb{;2)}
             + 9\,(5\K_{(4} +31\K_{(2}\K_2)\sp{\m\n}\Y_\m\sp\n\sb{;1)}\non\\
&&\qquad\quad
             + 4\,(6\K_{(5} +51\K_{(3}\K_2 +60\K_{(2}\K_3)\sp{\m\n}\Y_{\m)\n}  
\Big)
\ena
In (\ref{Zm1-5}) the parenthesis around the indices imply symmetrization 
only (but note that $\Zm_4$ and $\Zm_5$ appear in (\ref{a1-a5}) with at least
one respectively two contracted pair(s) of indices). All terms in (\ref{Zm1-5})
with explicit indices vanish upon full contraction. From (\ref{Zm2j}) we find
\bea\label{Zm(2)-(8)}
\Zm_{(2)}&=&\X_{(2)}+2  \Y_{(1}\Y_{1)}  \non\\
\Zm_{(4)}&=&\X_{(4)}+4\{\Y_{(1}\Y_{3)}\}+6\Y_{(2}\Y_{2)}+12\Y_{(1}\K_2\Y_{1)}
\non\\
\Zm_{(6)}&=&\X_{(6)}  
+ 6\{\Y_{(1}\Y_{5)}\}
+15\{\Y_{(2}\Y_{4)}\}
+24  \Y_{(3}\Y_{3)} 
+40  \Y_{(1}\K_4\Y_{1)} \non\\
&&
+66\{\Y_{(1}\K_3\Y_{2)}\}
+50\{\Y_{(1}\K_2\Y_{3)}\} 
+72  \Y_{(2}\K_2\Y_{2)} 
+153 \Y_{(1}\K_2\K_2\Y_{1)} \non\\
\Zm_{(8)}&=&\X_{(8)} 
+ 8\{\Y_{(1}\Y_{7)}\}
+28\{\Y_{(2}\Y_{6)}\} 
+56\{\Y_{(3}\Y_{5)}\}
+70  \Y_{(4}\Y_{4)} 
+84  \Y_{(1}\K_6\Y_{1)} \non\\
&&
+225  \{\Y_{(1}\K_5\Y_{2)}\}
+\sfrac{980}3 \{\Y_{(1}\K_4\Y_{3)}\} 
+273  \{\Y_{(1}\K_3\Y_{4)}\}
+126  \{\Y_{(1}\K_2\Y_{5)}\}  \non\\
&&
+480   \Y_{(2}\K_4\Y_{2)}
+520 \{\Y_{(2}\K_3\Y_{3)}\}
+288 \{\Y_{(2}\K_2\Y_{4)}\}
+\sfrac{1120}3\Y_{(3}\K_2\Y_{3)}  \non\\
&&
+\sfrac{6336}5 \Y_{(1}\K_3\K_3\Y_{1)}
+ 870 \{\Y_{(1}\K_2\K_4\Y_{1)}\}
+1440 \{\Y_{(1}\K_2\K_3\Y_{2)}\}  \non\\
&&
+\sfrac{6966}5 \{\Y_{(1}\K_3\K_2\Y_{2)}\}
+1092 \{\Y_{(1}\K_2\K_2\Y_{3)}\}
+\sfrac{7776}5 \Y_{(2}\K_2\K_2\Y_{2)} \non\\
&&
+ 3348 \Y_{(1}\K_2\K_2\K_2\Y_{1)}
\ena
The last term in $\Zm_{(8)}$ was obtained from (\ref{ZMloc}). From (\ref{ZSj})
we find that the scalar quantities are given explicitly by
\bea\label{Zs1-5}
\Zs_1 &=&\sfrac12\tr[\K_{1\m\m}] \non\\
\Zs_2 &=&\sfrac12\tr[\K_{2\m\m} +\sfrac65\K_2\K_{\m\m}]\,
-\sfrac12\tr[\K_{\m(1}]\tr[\K_{1)\m}]\, +3\K_{(2}\sp{\m\n}\tr[\K_{\m)\n}]\non\\
\Zs_3 &=&\sfrac12\tr[\K_{3\m\m} + 4\K_2\K_{1\m\m}]
-\sfrac32\tr[\K_{\m(1}] \tr[\K_{2)\m}]
+ 4\K_{(3}\sp{\m\n}\tr[\K_{\m)\n}] + 6\K_{(2}\sp{\m\n}\tr[\K_{1\m)\n}]   \non\\
\Zs_4 &=&\sfrac1{14}\tr\Big[7\K_{4\m\m} +40\K_2\K_{2\m\m} 
           +30\K_3\K_{1\m\m} +48\K_2\K_2\K_{\m\m}\Big]\ 
-\sfrac32\tr[\K_{\m(2}]\tr[\K_{2)\m}] \non\\ 
&&
-\sfrac25\tr[\K_{\m(1}]\tr[5\K_{3)\m} +6\K_2\K_{1)\m}]\ 
-  6 \tr[\K_{\m(1}]\,\K_2\sp{\m\n}\tr[\K_{1)\n}]\ 
+10\K_{(3}\sp{\m\n}\tr[\K_{1\m)\n}] \non\\
&&
+ (5\K_{(4} +36\K_{(2}\K_2)\sp{\m\n}\tr[\K_{\m)\n}] \ 
+ 2\K_{(2}\sp{\m\n}\tr[5\K_{2\m)\n}+6\K_2\K_{\m)\n}] \non\\
\Zs_5 &=&\sfrac12\tr\Big[\K_{5\m\m} +\sfrac{15}2\K_2\K_{3\m\m} 
      + 15\K_3\K_{2\m\m} + 48\K_2\K_2\K_{1\m\m} \Big] \non \\
&&
-\sfrac52\tr[\K_{\m(1}]\tr[\K_{4)\m} + 4\K_2\K_{2)\m}] \ 
-        \tr[\K_{\m(2}]\tr[5\K_{3)\m}+ 6\K_2\K_{1)\m}] \non\\
&&
- 30 \tr[\K_{\m(1}]\,\K_2\sp{\m\n}\tr[\K_{2)\n}]\ 
- 10 \tr[\K_{\m(1}]\,\K_3\sp{\m\n}\tr[\K_{1)\n}] \non\\
&&
+ 15 \K_{(2}\sp{\m\n}\tr[\K_{3\m)\n} + 4\K_{ 2}\K_{1\m)\n}]\ 
+  4 \K_{(3}\sp{\m\n}\tr[5\K_{2\m)\n} +6\K_{ 2}\K_{\m)\n}] \non\\
&&
+ 3\,(5\K_{(4} +36\K_{(2}\K_2)\sp{\m\n}\tr[\K_{1\m)\n}]\non\\
&&
+ 6\,( \K_{(5} +12\K_{(2}\K_3 +12\K_{(3}\K_2 )\sp{\m\n}\tr[\K_{\m)\n}]
\ena
where the parenthesis mean symmetrization only. Upon full contraction we find
from (\ref{Zs2j})
\bea\label{Zs(2)-(8)}
\Zs_{(2)} &=&\sfrac12\tr[\K_{(4)} +\sfrac65\K_{(2}\K_{2)}] \,
            -\sfrac12\tr[\K_{\m(1}]\tr[\K_{1)\m}] \non\\
\Zs_{(4)} &=&\sfrac1{14}\tr\Big[7\K_{(6)} +40\K_{(2}\K_{4)}
                        +30\K_{(3}\K_{3)} +48\K_{(2}\K_2\K_{2)} \Big] \,
-\sfrac32\tr[\K_{\m(2}]\tr[\K_{2)\m}]  \non\\
&&
- 2 \tr[\K_{\m(1}]\tr[\K_{3)\m} +\sfrac65\K_2\K_{1)\m}] \,
- 6 \tr[\K_{\m(1}]\,\K_2\sp{\m\n} \tr[\K_{1)\n}] \non\\
\Zs_{(6)} &=&\sfrac12\tr\Big[\K_{(8)} +\sfrac{28}3\K_{(2}\K_{6)} 
 + \sfrac{70}3\K_{(3}\K_{5)} +\sfrac{140}9\K_{(4}\K_{4)} 
 +\sfrac{272}3\K_{(2}\K_2\K_{4)} +\sfrac{400}3\K_{(2}\K_3\K_{3)} \non\\
&&
 +\sfrac{432}5 \K_{(2}\K_2\sp{2}\K_{2)}\Big]
-\sfrac37\tr[\K_{\m(1}]\tr\Big[7\K_{5)\m} +40\K_2\K_{3)\m}
   +30\K_3\K_{2)\m} +48\K_{2}\K_2\K_{1)\m}\Big]  \non\\
&&
-\sfrac{15}2\tr[\K_{\m(2}]\tr[\K_{4)\m} +4\K_2\K_{2)\m}]  \,
- \sfrac15\tr[5\K_{\m(3}+6\K_{\m(1}\K_2]\tr[5\K_{3)\m}+6\K_2\K_{1)\m}] \non\\
&&
- 3\tr[\K_{\m(1}]\, (5\K_4 +36\K_2\K_2)\sp{\m\n} \tr[\K_{1)\n}]\,
-12\tr[\K_{\m(1}]\,\K_2\sp{\m\n} \tr[5\K_{3)\n}+6\K_2\K_{1)\n}] \non\\
&&
-60\tr[\K_{\m(1}]\,\K_3\sp{\m\n} \tr[\K_{2)\n}] \,
-45\tr[\K_{\m(2}]\,\K_2\sp{\m\n} \tr[\K_{2)\n}]  \non\\
\Zs_{(8)} &=&\sfrac12\tr[\K_{(10)}] 
+\sfrac8{11}\tr\Big[9\K_{(2}\K_{8)} +500\K_{(2}\K_4\K_{4)} 
+1680\K_{(2}\K_2\sp{2}\K_{4)} +1296\K_{(2}\K_2\sp{3}\K_{2)} \Big] \non\\
&&
+\sfrac{21}{11}\tr\Big[12\K_{(3}\K_{7)} +24\K_{(4}\K_{6)} +15\K_{(5}\K_{5)}
               +\sfrac{304}5\K_{(2}\K_2\K_{6)} +290\K_{(2}\K_3\K_{5)} \non\\
&&\qquad\qquad
+ 460\K_{(2}\K_3\K_2\K_{3)} \Big]\ 
+ 540\tr[\K_{(3}\K_3\K_{4)}]
+1788\tr[\K_{(2}\K_2\K_3\K_{3)}] \non\\
&&
-4\tr[\K_{\m(1}]\tr\Big[\K_{7)\m} +\sfrac{28}3\K_2\K_{5)\m}
  +\sfrac{70}3\K_3\K_{4)\m} +\sfrac{140}9\K_4\K_{3)\m}
  +\sfrac{272}3\K_2\K_2\K_{3)\m} \non\\
&&\qquad\qquad\qquad\quad
  +\sfrac{400}3\K_2\K_3\K_{2)\m} +\sfrac{432}5\K_2\K_2\K_2\K_{1)\m}\Big] \non\\
&&
-14\tr[\K_{\m(2}]\tr\Big[\K_{6)\m} +\sfrac{15}2 \K_2\K_{4)\m}
                         +15\K_3\K_{3)\m} +48\K_2\K_2\K_{2)\m} \Big] \non\\
&&
-\sfrac45\tr[5\K_{\m(3} +6\K_{\m(1}\K_2]
     \tr[7\K_{5)\m} +40\K_2\K_{3)\m} +30\K_3\K_{2)\m} +48\K_2\K_2\K_{1)\m}]\non\\
&&
-\sfrac{35}2\tr[\K_{\m(4}+4\K_{\m(2}\K_2]\tr[\K_{4)\m} +4\K_2\K_{2)\m}] \non\\
&&
-  4\tr[\K_{\m(1}]\, (7\K_6 +125\{\K_2\K_4\} +170\K_3\K_3
                    +720\K_2\K_2\K_2 )\sp{\m\n} \tr[\K_{1)\n}] \non\\
&&
-168\tr[\K_{\m(1}]\,  (\K_5 + 12\{\K_2\K_3\})\sp{\m\n} \tr[\K_{2)\n}] \non\\
&&
-\sfrac{56}5\tr[\K_{\m(1}]\, (5\K_4 +36\K_2\K_2)\sp{\m\n} 
    \tr[5\K_{3)\n}+6\K_2\K_{1)\n}] \non\\
&&
-280\tr[\K_{\m(1}]\,\K_3\sp{\m\n} \tr[\K_{4)\n} +4\K_2\K_{2)\n}]\,
- 42\tr[\K_{\m(2}]\, (5\K_4 +36\K_2\K_2)\sp{\m\n} \tr[\K_{2)\n}] \non\\
&&
- 24\tr[\K_{\m(1}]\,\K_2\sp{\m\n} \tr[7\K_{5)\n} +40\K_2\K_{3)\n}
                +30\K_3\K_{2)\n} +48\K_2\K_2\K_{1)\n} ] \non\\
&&
-112\tr[\K_{\m(2}]\,\K_3\sp{\m\n} \tr[5\K_{3)\n} +6\K_2\K_{1)\n}]\ 
-420\tr[\K_{\m(2}]\,\K_2\sp{\m\n} \tr[ \K_{4)\n} +4\K_2\K_{2)\n}] \non\\
&&
-\sfrac{56}5\tr[5\K_{\m(3} +6\K_{\m(1}\K_2]\,\K_2\sp{\m\n}
          \tr[5\K_{3)\n} +6\K_2\K_{1)\n}] 
\ena
where the seven terms of fifth order were found from (\ref{ZSloc}) and also 
from expanding $B$ through tenth order in the normal coordinates (see appendix
E).

To return to conventional notation we use
\bea
\Z       &=& X +\sfrac16 R \non\\
\Z_\m    &=& X_{;\m}  +\sfrac13 F_\m\sp\n\sb{;\n}+\sfrac16 R_{;\m} \\
\Z_{(2)} &=& X_;\sp\m\sb\m + \sfrac12 F^{\m\n} F_{\m\n}
               +\sfrac15 R_;\sp\m\sb\m + \sfrac1{12} R^2 
 -\sfrac1{30} R^{\m\n} R_{\m\n} +\sfrac1{30} R^{\k\l\m\n} R_{\k\l\m\n}\non 
\ena
etc (these suffice for $\sa_1$ and $\sa_2$). In $\sa_4$ respectively $\sa_5$ 
we have e.g.
\bea
\Z_{(1}\do\HZ\Z_{1)}&=& 
\Z_\m\do (\Z g^{\m\n} +F^{\m\n} -\sfrac13 R^{\m\n})\Z_\n \\
\Z_{(2}\do\HZ\Z_{2)}&=& \sfrac13 \Z_{(2)}\Z\Z_{(2)} 
+\sfrac23\Z_{\k\m}\do (\Z g^{\m\n} +2F^{\m\n} -\sfrac23 R^{\m\n})\Z_{\k\n} 
+\sfrac49\Z_{\k\m}\do R^{\k\l\m\n}\Z_{\l\n}  \non
\ena
Using such translations we have verified that our results for the first four 
coefficients agree with earlier authors, in particular \cite{DeW}, \cite{Gi2} 
and \cite{Avr}, after accounting for differences in conventions and the 
occasional typographical mistake.

Finally, to illustrate the compactness of our notation, consider the special 
case of a scalar field in a Ricci flat space. Then only the Weyl tensor and 
its covariant derivatives can appear in the heat kernel coefficients. 
From appendix D of \cite{FKW} we can read off the total number of general 
coordinate scalars in that case. Thus, for $j=2$, 3 and 4 we expect 1, 3 and 12
terms in $\sa_j$, respectively. In our notation, see (\ref{trKj}), 
Ricci-flatness implies that $\tr[\K_j]$ vanishes. Only a few terms remain then
in (\ref{ZSj}, \ref{Zs2j}). In particular, $\Z$ and $\Z_1$ vanish and $\sa_2$ 
through $\sa_4$ contain only 
\bea
\sa_2 & : & \tr[\K_{(2}\K_{2)}] \non\\
\sa_3 & : & \tr[\K_{(2}\K_{4)}]\ ,\ \tr[\K_{(3}\K_{3)}]
       \ ,\ \tr[\K_{(2}\K_2\K_{2)}] \non\\
\sa_4 & : & \tr[\K_{(2}\K_{6)}]\ ,\ \tr[\K_{(3}\K_{5)}]\ ,\ \tr[\K_{(4}\K_{4)}]
       \ ,\ \tr[\K_{(2}\K_2\K_{4)}]\ ,\ \tr[\K_{(2}\K_3\K_{3)}]\ ,\\
&\phantom{:}& 
            \tr[\K_{(2}\K_2\K_2\K_{2)}]\ ,\ 
\tr[\K_{\m(1}\K^\m\sb{1}]\tr[\K^\n\sb{1}\K_{1)\n}]\ ,\
\tr[\K_{\m(1}\K_2]\tr[\K_2\K_{1)\m}]\ ,\ (\tr[\K_{(2}\K_{2)}])^2 \non
\ena
Here there are 1, 3 and 9 terms respectively, so that, starting with $\sa_4$, 
our notation is not only index-free, but also generates less terms. 
This becomes even more pronounced for $\sa_5$ where \cite{FKW} informs us 
that in the Ricci-flat case there are 67 terms, however in our notation there 
are only 17 terms.
\section{Conclusions}
\setcounter{equation}{0}
We have presented the explicit diagonal values of the first five (six) heat 
kernel coefficients for a general Laplace-type operator on a Riemannian 
(respectively flat) space.  To solve the pertinent recursion relations, we 
relied not only on well known techniques (matrix notation for field indices, 
Fock-Schwinger gauge and Riemann normal coordinates), but also used a new 
notation free of spacetime indices. It is this latter compact notation which 
allows us to write down the fifth and sixth coefficient in the first place. 
Insisting also on manifest hermiticity of the results, the fifth (sixth) 
coefficient has 26 (respectively 75) terms. They were presented here for the 
first time. Beyond these coefficients, the leading terms -- to the same order 
in derivatives or curvatures as needed for the fourth coefficient -- for any 
heat kernel coefficient were given in the general case. To determine the heat 
kernel coefficients, we have found it useful to proceed in a few steps, 
starting from the simplest case of a flat spacetime without a gauge field. 
We could show that `turning on a gauge field' and next `curving spacetime' is 
taken care of by specific covariant substitutions in the flat space 
coefficients. With this `dressing up' understood, the number of terms and their
numerical prefactors do not change in the process.

In our notation, a typical term in a heat kernel coefficient consists of a 
maximally symmetrized (product of) covariant derivatives of the basic 
curvatures, made into a scalar by contracting all indices. Consider e.g. the 
following term
\bee\label{K3K3K4}
\tr[\K_{(3}\K_3\K_{4)}] \= \sfrac3{20} R^\a\sb{(\k}\sp\b\sb{\k;\l}
R^\b\sb\l\sp\g\sb{\m;\m} R^\g\sb\n\sp\a\sb{\n;\r\r)}
\ene
which appears in the fifth coefficient. Since we do not integrate over 
spacetime, partial integrations are not permitted. Furthermore, as long as we
do not write out the symmetrization, the Bianchi identity can not be used here.
Thus our notation gives a certain degree of uniqueness to the appearance of the
heat kernel coefficients, missing in a more conventional notation with 
explicit spacetime indices.

Of course, depending on the application one has in mind, our notation may or 
may not be useful. We plan to calculate the chiral anomaly based on our 
results, i.e. essentially evaluate the spinor-trace $\tr[\g_{2j+1}\sa_j]$ in 
$d=2j$ dimensions.
In this case elegant and complete results are known \cite{ZWZ}, using the 
language of differential forms. It should be interesting to see how this can 
be related to our use of symmetric tensors. Also the gravitational anomalies 
\cite{AGW} in $d=4j+2$ are computable in our framework. Possibly, the absence 
of certain terms from the heat kernel coefficients found here, see 
(\ref{overlap}), plays a role in this connection. 

Another task, now in progress \cite{BoV}, is to work out the functional trace 
of the diagonal heat kernel coefficients given here. In that case we can 
maintain our index-free notation (e.g. the integral of (\ref{K3K3K4}) is easily
seen to vanish without writing out the indices).
Although it would probably hold few surprises, {\it cf} \cite{FrT}, the 
result for the integrated and traced fifth coefficient could be used to study 
for the first time the one-loop short distance divergences of ten-dimensional 
supergravity.

As this paper neared completion, we were informed by the authors of 
\cite{FHSS1} that they had extended their results so as to include gauge 
fields. In \cite{FHSS2} they present the functional trace of the first six 
heat kernel coefficients for this case in flat spacetime, reduced to a 
so-called minimal basis. It should be possible to compare their result to the
sixth heat kernel coefficient as presented here, after taking its trace and 
integrating it. 

Finally, to avoid misunderstanding, we should mention that a recent preprint 
\cite{Kir} with the title ``The $a_5$ heat kernel coefficient on a manifold 
with boundary'' is not concerned with the fifth heat kernel coefficient as
presented here. Rather, on a manifold with boundary the expansion (\ref{hkc})
is in powers of $\sqrt\t$ rather than $\t$ and therefore our $a_j$ corresponds
to $a_{2j}$ of \cite{Kir}. Thus, in the terminology of \cite{Kir} we have 
determined here `the volume part of $a_{10}$'. 
\vskip.5cm
I would like to thank J.-P. B\"ornsen for his assistance in using Mathematica
to verify some of the results presented here. I am also grateful to I. Avramidi
for useful correspondence.
\newpage
\newpage
\begin{appendix}
\section{Notation}
\setcounter{section}{1}
\setcounter{equation}{0}
\renewcommand{\theequation}{\Alph{section}.\arabic{equation}}
We use Greek letters for spacetime indices. Indices $j,\,k,\,\el,\dots$ 
simply enumerate various objects. We take differential operators to act on 
everything to their right. If this is not intended, we use comma (semicolon) 
notation for partial (covariant) derivatives
\bee
F_{,\m} \,\equiv\, [\pa_\m, F] \quad ,\quad F_{;\m} \,\equiv\,[\de_\m, F]
\ene
A semicolon denotes simultaneous gauge and gravitational covariant 
differentiation. We use $a\equiv b$ to define $a$ in terms of $b$.
Our curvature conventions are fixed by ($V_\l$ is a gauge singlet)
\bee\label{Rcon}
[\de_\n,\de_\m] V_\l \= 2\,V_{\l;[\m\n]}\= R^\k\sb{\l\m\n} V_\k \quad ,\quad
R_{\l\n}\,\equiv\,R^\k\sb{\l\k\n} \quad ,\quad R\,\equiv\, R^\l\sb\l
\ene
In Fock-Schwinger gauge and Riemannian coordinates, see (\ref{FSgauge}) and
(\ref{Rnc}), we have\footnote{\mbox{Parenthesis around $j$ explicit indices 
denotes total symmetrization (with division by $j!$).}}
\bee
F_{,\m_1\dots\m_j}(0) = F_{;(\m_1\dots\m_j)}(0) \quad ,
\ene
for any, possibly matrix valued, general coordinate scalar $F(x)$. This 
property allows one to immediately covariantize the partial derivatives. Due 
to the total symmetry here it is convenient and sufficient to keep track of 
only the number of indices, leading us to define the {\it sans serif} symbols
\bee
\F_j\= F_{;(\m_1\dots\m_j)}(0)
\ene
We thus use an index-free notation for totally symmetrized tensors.
At the same time, such a symbol implies evaluation at the origin.
Following Avramidi, we use parenthesis around an even enumerative label to 
indicate not only total symmetrization, but also full contraction as in
\bee
\F_{(2j)} \= F_{;(\m_1\m_1\dots\m_j\m_j)}(0)
\ene
Since the metric at the origin is flat there is no need to write the indices
in their covariant respectively contravariant positions.
{\it Par abus de language} an odd label enclosed in parenthesis denotes 
symmetrization and contraction of all but one of the involved indices. This 
will occur only if there is exactly one other factor with an odd label,
as e.g. in
\bee
\F_{(3)} \G_{(3)}\= F_{;(\k\m\m)}(0) G_{;(\k\n\n)}(0)
\ene
We generalize Avramidi's notation by allowing such a simultaneous 
symmetrization and contraction to extend over several factors, e.g.
\bee\label{FkGl}
\F_{(k}\G_{2j-k)}\=F_{;(\m_1\,\dots\,\m_k}(0)\,
G_{;\m_{k+1}\,\dots\,\m_{2j})}(0)
\,\d^{\m_1\m_2}\,\dots\,\d^{\m_{2j-1}\m_{2j}}
\ene
If desired, such an expression can be written out in such a way that the total 
symmetrizations extend only over the individual factors, as e.g. in
\bee
\F_{(2}\G_{2)} \= \sfrac13 F_{;(\m\m)}(0) G_{;(\n\n)}(0)
+\sfrac23 F_{;(\m\n)}(0) G_{;(\m\n)}(0)
\ene
In general, expanding (\ref{FkGl}) in this way yields at first $(2j-1)!!$ 
terms. However, due to the total symmetry of the covariant derivatives acting 
on each factor and assuming that $k\leq j$, there remain only $[k/2]+1$ terms, 
$[k/2]$ being the maximal number of self-contractions possible for $\F_k$. 
Finding the coefficient of each term is a combinatorical problem with the 
following solution
\bea\label{combi}
\F_{(2j}\G_{2k)} &=& {1\ov (2j+2k-1)!!} \su_{\el=0}^j 
P^{2j}_{j-\el} P^{2k}_{k-\el} (2\el)! \,
\F^{(2j-2\el)}_{\m_1\dots\m_{2\el}}\,\G^{(2k-2\el)}_{\m_1\dots\m_{2\el}}\non\\
\F_{(2j+1}\G_{2k+1)} &=& {1\ov (2j+2k+1)!!} \su_{\el=0}^j 
P^{2j+1}_{j-\el} P^{2k+1}_{k-\el} (2\el +1)!\,
\F^{(2j-2\el)}_{\m_1\dots\m_{2\el+1}}\,\G^{(2k-2\el)}_{\m_1\dots\m_{2\el+1}}
\non\\
P^j_k &\equiv & {j\ch 2k}(2k-1)!!\quad ,\quad k\leq [j/2]\quad ,\quad 
(-1)!! \equiv 1
\ena
Here $P^j_k$ is the number of ways in which one can choose $k$ {\it pairs} out
of $j$ objects. The covariant derivatives on $\F$ respectively $\G$ on the 
right hand side are understood to have been totally symmetrized. In general
there will be more than two factors, but in practice these are easily taken 
care of, as e.g. in
\bee
\F_{(2}\G_2\H_{2)} \= \sfrac1{15} \F_{(2)}\G_{(2)}\H_{(2)}
+ \sfrac2{15}(\F_{(2)}\G_{;(\m\n)}\H_{;(\m\n)} + 2\ {\rm more})
+ \sfrac8{15}\F_{;(\m\n)}\G_{;(\n\r)}\H_{;(\r\m)} 
\ene
Multinomial coefficients are defined as usual by
\bee
{j\ch k_1,\dots,k_N}\= {j!\ov k_1 !\dots k_{N+1}!}\quad ,\quad
k_{N+1}\,\equiv\,j\,-\su_{n=1}^N k_n
\ene
If the index $k_{N+1}$ appears in the summand of a sum involving this 
multinomial coefficient, then it is understood to have the indicated value. 
This convention applies also to alphabetic ordering as e.g. in
\bee\label{alphaord}
\su_{k=0}^j {j\ch k} F_k G_\el \,\equiv\, \su_{k=0}^j {j\ch k} F_k G_{j-k}
\ene
\section{Moving boxes}
\setcounter{equation}{0}
In the main text we gave an expression, eq (\ref{Xres}), for the diagonal 
values of the heat kernel coefficients $\sa_j$ through third order in the 
matrix potential $\X$ which holds only for $j\leq 5$. We indicated in 
(\ref{mobo}) how to obtain $\sa_6$ as well without generating terms with 
overlapping derivatives. In the general case we rely on the following lemma to
move the boxes from $G$ to $F$
\bea
F_{(2n} G_{2p)(2q-2p)}&=&\su_{j=0}^n \b_j(n,p,q) F_{(2n-2j)(2j} G_{2q)} 
\quad ,\quad n\leq p\leq q  \\
\b_j(n,p,q) &=& {q-p\ch n-j}\ {f(j,p,q)\ov f(n,q,p)} \quad ,\quad
f(j,p,q) \= (2j+2q-1)!!\, P^{2p}_{p-j} \non
\ena
with a similar expression for the case $F_{(2n+1} G_{2p+1)(2q-2p-2)}$.
Note that this can only be done when all indices are contracted. 
To prove the lemma, expand both sides using (\ref{combi}) and compare.
We also note the following exceptional case
\bee
F_{(k+2n} G_{k)(2n)} \= F_{(2n)(k} G_{k+2n)} 
\ene
\section{$\sa_6$ in flat space}
\setcounter{equation}{0}
Expressions (\ref{deriv}), (\ref{Xres}), (\ref{Subs}) and (\ref{Zres}) 
suffice to write down the sixth heat kernel coefficient in a flat space, but
with a gauge connection. We obtain
\bea
&&\sa_6 =\Z^6
+\sfrac57\{\Z^4\Z_{(2)}\} 
+\sfrac87\{\Z^3\Z_{(2)}\Z\}
+\sfrac97\Z^2\Z_{(2)}\Z^2 
+\sfrac27\Z_{(1}\do\HZ^3\Z_{1)} 
+\sfrac47\{\Z\Z_{(1}\do\HZ^2\Z_{1)}\} \non\\
&&
+\sfrac67\{\Z^2\Z_{(1}\do\HZ\Z_{1)}\}
+\sfrac87\Z\Z_{(1}\do\HZ\Z_{1)}\Z 
+\sfrac87\{\Z^3\Z_{(1}\do\Z_{1)}\} 
+\sfrac{12}7\{\Z^2\Z_{(1}\do\Z_{1)}\Z\} 
+\sfrac5{14}\{\Z^3\Z_{(4)}\} \non\\
&&
+\sfrac9{14}\{\Z^2\Z_{(4)}\Z\}  
+\sfrac{12}7\{\Z\Z_{(1}\do\Z_{3)}\Z\}
+\sfrac{15}{14}\{\Z^2\Z_{(1}\do\Z_{3)}\} 
+\sfrac97\{\Z^2\Z_{(3}\do\Z_{1)}\} 
+\sfrac57\{\Z\Z_{(1}\do\HZ\Z_{3)}\} \non\\
&&
+\sfrac67\{\Z\Z_{(3}\do\HZ\Z_{1)}\} 
+\sfrac5{14}\{\Z_{(1}\do\HZ^2\Z_{3)}\}
+\sfrac97\{\Z^2\Z_{(2}\do\Z_{2)}\}
+\sfrac37\{\Z^2\Z_{(2)}\sp{2}\}
+\sfrac3{14}\Z_{(2}\do\HZ^2\Z_{2)} \non\\
&&
+\sfrac37 \Z_{(2)}\Z^2\Z_{(2)} 
+\sfrac9{14}\{\Z\Z_{(2}\do\HZ\Z_{2)}\} 
+\sfrac47\{\Z\Z_{(2)}\Z\Z_{(2)}\} 
+\sfrac{27}{14}\Z\Z_{(2}\do\Z_{2)}\Z
+\sfrac47 \Z\Z_{(2)}\sp{2}\Z \non\\
&&
+\sfrac9{14}\{\Z_{(1}\do\HZ_1\HZ\Z_{2)}\} 
+\sfrac47\{\Z_{(1}\do\Z_{1)}\Z\Z_{(2)}\}   
+\sfrac{12}7\{\Z\Z_{(1}\do\HZ_1\Z_{2)}\} 
+\sfrac87\{\Z\Z_{(1}\do\Z_{1)}\Z_{(2)}\}  \non\\
&&
+\sfrac{27}{14}\{\Z\Z_{(2}\do\HZ_1\Z_{1)}\} 
+\sfrac47\{\Z\Z_{(2)}\Z_{(1}\do\Z_{1)}\} 
+\sfrac67\{\Z_{(1}\do\HZ\HZ_1\Z_{2)}\} 
+\sfrac27\{\Z_{(1}\do\HZ\Z_{1)}\Z_{(2)}\}  \non\\
&&
+\sfrac{18}7\{\Z\Z_{(1}\do\HZ_2\Z_{1)}\}  
+\sfrac97\{\Z_{(1}\do\HZ\HZ_2\Z_{1)}\} 
+\sfrac{27}{14}\Z_{(1}\do\HZ_1\HZ_1\Z_{1)}  
+\sfrac47\Z_{(1}\do\Z_{1)}\,\Z_{(1}\do\Z_{1)} \non\\
&&
+\sfrac5{42}\{\Z^2\Z_{(6)}\} 
+\sfrac4{21}\Z\Z_{(6)}\Z  
+\sfrac5{21}\{\Z_{(1}\do\HZ\Z_{5)}\}
+\sfrac{10}{21}\{\Z\Z_{(1}\do\Z_{5)}\} 
+\sfrac47 \{\Z\Z_{(5}\do\Z_{1)}\} \non\\
&&
+\sfrac{25}{84}\{\Z_{(2}\do\HZ\Z_{4)}\}  
+\sfrac{25}{28}\{\Z\Z_{(2}\do\Z_{4)}\} 
+\sfrac{20}{21}\{\Z\Z_{(4}\do\Z_{2)}\} 
+\sfrac{20}{21}\{\Z\Z_{(3}\do\Z_{3)}\} 
+\sfrac5{21}\Z_{(3}\do\HZ\Z_{3)} \non\\
&&
+\sfrac{25}{28}\{\Z_{(1}\do\HZ_1\Z_{4)}\}  
+\sfrac{10}7  \Z_{(1}\do\HZ_4\Z_{1)} 
+\sfrac{10}7\{\Z_{(1}\do\HZ_2\Z_{3)}\} 
+\sfrac{40}{21}\{\Z_{(1}\do\HZ_3\Z_{2)}\} 
+\sfrac{20}{21}\{\Z_{(2}\do\HZ_1\Z_{3)}\} \non\\
&&
+\sfrac{40}{21} \Z_{(2}\do\HZ_2\Z_{2)} 
+\sfrac3{14}\{\Z\Z_{(4)}\Z_{(2)}\} 
+\sfrac5{28}\{\Z_{(2)}\Z\Z_{(4)}\} 
+\sfrac5{28}\{\Z\Z_{(2)}\Z_{(4)}\}
+\sfrac37\{\Z\Z_{(3)}\do\Z_{(3)}\} \non\\
&&
+\sfrac27\Z_{(3)}\do\HZ\Z_{(3)}
+\sfrac5{28}\{\Z_{(1}\do\Z_{1)}\Z_{(4)}\}  
+\sfrac9{14}\{\Z_{(1}\do\Z_{2)}\Z_{(3)}\} 
+\sfrac37\{\Z_{(2}\do\Z_{1)}\Z_{(3)}\} 
+\sfrac5{14}\{\Z_{(2)}\Z_{(1}\do\Z_{3)}\} \non\\
&&
+\sfrac37\{\Z_{(1}\do\Z_{3)}\Z_{(2)}\} 
+\sfrac37\{\Z_{(2}\Z_{2)}\Z_{(2)}\}  
+\sfrac17  \Z_{(2)}\sp{3} 
+\sfrac1{42}\{\Z\Z_{(8)}\} 
+\sfrac2{21}\{\Z_{(1}\do\Z_{7)}\}
+\sfrac29\{\Z_{(2}\do\Z_{6)}\} \non\\
&&
+\sfrac13\{\Z_{(3}\do\Z_{5)}\} 
+\sfrac13\Z_{(4}\do\Z_{4)} 
+\sfrac5{126}\{\Z_{(2)}\Z_{(6)}\} 
+\sfrac5{42}\{\Z_{(3)}\do\Z_{(5)}\} 
+\sfrac3{14}\Z_{(2)(2}\do\Z_{2)(2)} 
+\sfrac1{462} \Z_{(10)}\non\\
\ena
The $\Z$'s appearing here were defined in (\ref{Zj}), (\ref{Z2j}) and 
(\ref{Zhatdef}). Note in particular that $\Z\=\X$. There is a total of 75 
terms. In curved space, $\sa_6$ will look exactly the same, but in that case 
we do not know the values for some of the $\Z$'s.
\newpage
\section{Consequences of Fock-Schwinger gauge}
\setcounter{equation}{0}
To prove (\ref{AF}), we start from
\bee
(A_{(\m_1} F)_{,\m_2\dots\m_j)}(0) \= 0
\ene
which holds for any function $F$ in Fock-Schwinger gauge, see (\ref{symA}).
Now take $j=2n+2$, contract {\it all} indices and write out with respect to
the index of the gauge connection. It is essential for the proof that, due to 
the full contraction, all indices here are dummies. 

A further consequence of (\ref{symA}) is the following.
If we define $\check\Z$ to act to its {\it left} as follows
\bee
\Z_{(j}\do{\check\Z}_n\do\,\equiv\,
\Z_{(j}\do \Z_n\do \,-\,{2j\ov n+1}\Z_{\n(j-1}\do\Y_{n+1}\sp\n 
\ene
where we left out inessential factors to the right, then we have the following
alternative notations
\bea
\Z_{(j}\do{\check\Z}_n\do\Z_{k)} &=& \Z_{(j}\do\HZ_n\Z_{k)} \non\\
\Z_{(j}\do{\check\Z}_p\do\HZ_n\Z_{k)} &=& \Z_{(j}\do\HZ_p\HZ_n\Z_{k)}
\ena
In the last case the `check' notation has the advantage that it generates less
terms than the `hat' notation, namely
\bea
\Z_{(j}\do{\check\Z}_p\do\HZ_n\Z_{k)} &=& \Z_{(j}\do\Z_p\do\Z_n\Z_{k)} 
\,-\,{2j\ov p+1}\,\Z_{\m(j-1}\do\Y_{p+1}\sp\m\Z_n\Z_{k)} 
\,+\,{2k\ov n+1}\,\Z_{(j}\do\Z_p\do\Y_{n+1}\sp\n\Z_{k-1)\n} \non\\
&&\quad
\,-\,{4jk\ov (p+1)(n+1)}\,\Z_{\m(j-1}\dg\Y_{p+1}\sp\m\Y_{n+1}\sp\n\Z_{k-1)\n}
\ena
%
% (A generalization of this lemma is given by 
% $$\F_{(2j-k}(\HX\sa)_{k)} \= \F_{(2j-k} (\Z\dg\sa)_{k)}\, 
% - \,{2j-k\ov k+1}\,\F_{\m (2j-k-1} (2\A^\m\sa)_{k+1)}$$
% which we can use to eliminate $Z(A\cd\pa)^2Z$.)
%
\section{How to get your $\Z$'s}
\setcounter{equation}{0}
Here we shall give some details on how to obtain (\ref{ZSj}). The steps for 
(\ref{Zmj}) are very similar so we omit them. Our starting point is 
(\ref{Zscalar}). We take $j$ partial derivatives of this expression and 
evaluate at the origin ($m\equiv j-k-\el$) to obtain
\bee
\Zs_j = \sfrac12\su_{k=0}^j {j+1\ch k} \sg^{\m\n}\sb{(k}\B_{j-k\,\m)\n}
-\sfrac14\su_{k=1}^{j-1}\su_{\el=1}^{j-k} {j\ch k,\el}
\B_{\m(k}\sg^{\m\n}\sb{m}\B_{\el)\n} 
\ene
where we used that $B$ and its first derivative vanish there.
Separate off the terms with undifferentiated inverse metric
\bea\label{diffm}
\Zs_j &=& \sfrac12\,\B_{\m\m\,j}
+\sfrac12\su_{k=2}^j {j+1\ch k}\sg^{\m\n}\sb{(k}\B_{j-k\,\m)\n} \non\\
&&\quad
-\sfrac14\su_{k=1}^{j-1} {j\ch k} \B_{\m(k} \B_{\el)\m}
-\sfrac14\su_{k=1}^{j-3} \su_{\el=1}^{j-k-2} {j\ch k,\el}
                                \B_{\m(k}\sg^{\m\n}\sb{m}\B_{\el)\n} 
\ena
Inserting (\ref{invmet}) and (\ref{Bj}) and collecting terms of the same order
in the curvatures $\K$ and $\Y$, we obtain (\ref{ZSj}). Up to here all 
contractions were explicitly indicated and the parenthesis denoted 
symmetrization only. If we now replace $j$ by $2j$ and fully contract,
the second term in (\ref{diffm}) vanishes due to (\ref{gsym0}) and we find
\bee
\Zs_{(2j)} = \sfrac12\,\B_{(2j+2)}
-\sfrac14\su_{k=1}^{2j-1} {2j\ch k} \B_{\m(k} \B_{\el)\m}
-\sfrac14\su_{k=1}^{2j-3} \su_{\el=1}^{2j-k-2} {2j\ch k,\el}
    \B_{\m(k}\sg^{\m\n}\sb{m}\B_{\el)\n} 
\ene
This yields (\ref{Zs2j}). Thus we see that, since $B$ is at least of first 
order in curvature, finding the heat kernel coefficients through order $n$ 
in the curvature requires the inverse metric only through order $n-2$. 

Writing $G$ for the inverse metric, we find through sixth order in the 
normal coordinates
\bea\label{invmetexpl}
&& G(x) \= \I +\,2\K_2 {x^2\ov 2!}\,+\,
2\K_3 {x^3\ov 3!}\,+\,
2(\K_4 +\sfrac{36}{5}\K_2\sp{2}) {x^4\ov 4!}\,+\,
2(\K_5 + 12\{\K_2\K_3\}) {x^5\ov 5!} \non\\
&&\quad 
+2(\K_6 +\sfrac{125}7 \{\K_2\K_4\} +\sfrac{170}7 \K_3\sp{2}
   +\sfrac{720}7 \K_2\sp{3}) {x^6\ov 6!}\,+\, O(x^7) 
\ena
For the logarithm of the Van Vleck-Morette determinant we find through tenth 
order in the normal coordinates
\bea\label{Bexpl}
&&B(x) \=\tr\Big[\,
 \K_2 {x^2\ov 2!}\,+\, 
 \K_3 {x^3\ov 3!}\,+\,
(\K_4 +\sfrac65\K_2\sp{2}) {x^4\ov 4!}\,+\,
(\K_5 + 4 \K_2\K_3  )      {x^5\ov 5!}      \non\\
&&
+(\K_6+\sfrac{40}7\K_2\K_4+\sfrac{30}7\K_3\sp{2}
      +\sfrac{48}7\K_2\sp{3}){x^6\ov 6!}\,
+\,(\K_7+\sfrac{15}2\K_2\K_5 + 15\K_3\K_4 + 48\K_2\sp{2}\K_3) {x^7\ov 7!}\non\\
&&
+(\K_8+\sfrac{28}3\K_2\K_6 +\sfrac{70}3\K_3\K_5+\sfrac{140}9\K_4\sp{2}
      +\sfrac{272}3\K_2\sp{2}\K_4 +\sfrac{400}3\K_2\K_3\sp{2}
      +\sfrac{432}5\K_2\sp{4}) {x^8\ov 8!} \non\\
&&
+(\K_9
+\sfrac{ 56}5\K_2\K_7
+\sfrac{168}5\K_3\K_6 
+ 56\K_4\K_5 
+\sfrac{756}5\K_2\sp{2}\K_5 
+ 584\K_2\K_3\K_4 
+ 144\K_3\sp{3} \non\\
&&\qquad 
+\sfrac{5184}5\K_2\sp{3}\K_3) {x^9\ov 9!}\non\\
&&
+(\K_{10}
+\sfrac{144}{11}\K_2\K_8   +\sfrac{504}{11}\K_3\K_7 +\sfrac{1008}{11}\K_4\K_6 
+\sfrac{630}{11}\K_5\sp{2} +\sfrac{12768}{55}\K_2\sp{2}\K_6 
+\sfrac{12180}{11}\K_2\K_3\K_5 \non\\
&&\qquad
+\sfrac{8000}{11}\K_2\K_4\sp{2} +1080\K_3\sp{2}\K_4
+\sfrac{26880}{11}\K_2\sp{3}\K_4 +3576\K_2\sp{2}\K_3\sp{2} 
+\sfrac{19320}{11}\K_2\K_3\K_2\K_3 \non\\
&&\qquad
+\sfrac{20736}{11}\K_2\sp{5}) {x^{10}\ov 10!}\,\Big]\ +\ O(x^{11})
\ena
%
% IF we treat K_j as commuting, the total coeff for K2^2K3^3 is 58656/11.
%
For the gauge connection (with covariant index) it suffices to know that
\bea\label{Aexpl}
A(x) \! &=&\!  \Y_1 x^1\,  +\Y_2 {x^2\ov 2!}\,
+(\Y_3 -{3\ov 2}\K_2\Y_1) {x^3\ov 3!}\,
+(\Y_4 -{18\ov 5}\K_2\Y_2 -{8\ov 5}\K_3\Y_1) {x^4\ov 4!} \non\\
&&\quad
+\Big(\Y_5 -{20\ov 3}\K_2\Y_3 - 5\K_3\Y_2 -\sfrac13(5\K_4-9\K_2\sp{2})\Y_1\Big)
 {x^5\ov 5!} \non\\
&&\quad
+\Big(\Y_6-{75\ov 7}\K_2\Y_4-{80\ov 7}\K_3\Y_3-{9\ov 7}(5\K_4-9\K_2\sp{2})\Y_2
\non\\  
&&\quad\qquad  -{12\ov 7}(\K_5-2\K_3\K_2-4\K_2\K_3)\Y_1\Big) {x^6\ov 6!} \ 
+\ O(x^7)
\ena
To convert the above expansions to conventional notation, use (\ref{defY}) 
and (\ref{defK}). 
\section{Symmetry factors}
\setcounter{equation}{0}
Below we give a table containing all symmetry factors $S/P_N$, defined in 
(\ref{symmfac}), for $N\leq 4$. The list $k_1\dots k_N$ in the 
second column is an abbreviation for $\tr[\K_{k_1}\dots \K_{k_N}]$. 
It is to be understood that $k$, $\el$, $m$ and $n$ are different integers,
each having at least the value 2.
\vskip.5cm
\begin{center}
\begin{tabular}{|c|c|c|c|}\hline
$N$ & $k_1\dots k_N$    & $S$ & $S/P_N$ \\ \hline\hline 
    & $k\ k$            &  1  &  1/2    \\ \cline{2-4}
\raisebox{1.5ex}[-1.5ex]{2}
    & $k\ \el$            &  2  &   1     \\ \hline
    & $k\ k\ k$         &  1  &  1/6    \\ \cline{2-4}
  3 & $k\ k\ \el$       &  3  &  1/2    \\ \cline{2-4}
    & $k\ \el\ m$       &  6  &   1     \\ \hline
    & $k\ k\ k\ k$      &  1  &  1/8    \\ \cline{2-4}
    & $k\ \el\ k\ \el$  &  2  &  1/4    \\ \cline{2-4}
    & $k\ k\ k\ \el$    &  4  &  1/2    \\ \cline{2-4}
  4 & $k\ k\ \el\ \el$  &  4  &  1/2    \\ \cline{2-4}
    & $k\ \el\ k\ m$    &  4  &  1/2    \\ \cline{2-4}
    & $k\ k\ \el\ m$    &  8  &   1     \\ \cline{2-4}
    & $k\ \el\ m\ n$    &  8  &   1     \\ \hline
\end{tabular}
\end{center}
\end{appendix}


\newpage
\begin{thebibliography}{99}
\bibitem{Fu1} S.~A. Fulling, editor, {\it Heat Kernel Techniques and Quantum 
Gravity}, Winnipeg, Canada 1994, published as vol. 4 of Discourses in 
Mathematics and Its Applications, Dept. of Math., Texas A \& M University, 
College Station, TX, 1995.
\bibitem{ABP} M.~F. Atiyah, R. Bott and V.~K. Patodi, Inv. Math. {\bf 19} 
(1973) 279.
\bibitem{Gi1} P.~B. Gilkey, {\it Invariance Theory, the Heat Equation and 
the Atiyah-Singer Index Theorem}, Math. Lect. Series vol. 11, Publish or 
Perish (1984).
\bibitem{Sch} J.~S. Schwinger, Phys. Rev. {\bf 82} (1951) 664.
\bibitem{DeW} B.~S. DeWitt, {\it Dynamical Theory of Groups and Fields}, 
(Gordon and Breach, New York, 1965, first published in {\it Relativity, Groups
and Topology}, Les Houches Summer School 1964); Phys. Rep. {\bf 19C} (1975) 
297.
%\bibitem{See} R.~T. Seeley, Proc. Symp. Pure Math.{\it in} Amer. Math. Soc. 
%{\bf 10} (1967) 288.
\bibitem{Sak} T.~Sakai, T\^ohoku Math. J. {\bf 23} (1971) 589.
\bibitem{Gi2} P.~B. Gilkey, J. Diff. Geom.  {\bf 10} (1975) 601;\ 
Compositio Math. {\bf 38} (1979) 201.
\bibitem{vdV} A.~E.~M. van de Ven, \np{250}{85}{593}.
\bibitem{Avr} I.~G. Avramidi, Teor. Mat. Fiz. {\bf 79} (1989) 219, 
(\tmp{79}{89}{494}); \np{355}{91}{712}; Ph~D thesis, Moscow State University, 
Moscow 1986, hep-th/9510140.
\bibitem{ABC} P. Amsterdamski, A.~L. Berkin and D.~J. O'Connor, 
\cqg{6}{89}{81}.
\bibitem{FHSS1} D. Fliegner, P. Haberl, M.~G. Schmidt, C. Schubert, Z. Phys. 
{\bf C64} (1994) 111; in [1], p. 87 (hep-th/9411177).
\bibitem{BELS} A.~A. Belkov, D. Ebert, A.~V. Lanyov and A. Schaale,
Int. J. Mod. Phys. {\bf C4} (1993) 775.\\
%\ Int. J. Mod. Phys. {\bf A8} (1993)  313.\\
A.~A. Belkov, A.~V. Lanyov and A. Schaale, Comp. Phys. Comm. {\bf 95} (1996) 
123.
\bibitem{AvS} I.~G. Avramidi and R.~Schimming, {\it in} Proceedings of the 
IIIrd Workshop ``Quantum Field Theory under the Influence of External 
Conditions'', Leipzig, 1995, hep-th/9510206.
\bibitem{BarV}  A.~O. Barvinsky and G.~A. Vilkovisky, \np{282}{87}{163};\ 
\np{333}{90}{471}, {\it idem} 512. 
\bibitem{FKW} S.~A. Fulling, R.~C. King, B.~G. Wybourne and C.~J. Cummins,
\cqg{9}{92}{1151}.
\bibitem{Bal} R.~D. Ball, Phys. Rep. {\bf 182} (1989) 1.
\bibitem{Ber} R.~A. Bertlmann, Anomalies in Quantum Field Theory, (Clarendon, 
1996).
%N.~K. Nielsen, Coincidence-limit method derivation of anomalies, 
%NORDITA preprint July 1978,\\ V.~N. Romanov and A.~S. Schwarz, Teor. Mat. 
%Fiz. {\bf 41} (1979) 190, (\tmp{41}{80}{967}),\\
%S.~M. Christensen and M.~J. Duff, \pl{76}{78}{571};\ \np{154}{79}{301}.
%J.~S. Dowker, J. Phys. {\bf A11} (1978) 347.
\bibitem{FrT} E.~S. Fradkin and A.~A. Tseytlin, \np{227}{83}{252}.
\bibitem{BaV} A.~O. Barvinsky and G.~A. Vilkovisky, 
Phys. Rep. {\bf 119} (1985) 1.
\bibitem{BGV} T.~P. Branson and P.~B. Gilkey, Com. Part. Diff. Eq. {\bf 15} 
(1990) 245.
\bibitem{McL} R.~G. McLenaghan, Proc. Camb. Phil. Soc. {\bf 65} (1969) 139;\ 
Ann. Inst. Henri Poincar\'e, {\bf XX} (1974) 153.
\bibitem{Par} L.~Parker, {\it in} ``Recent Developments in Gravitation'', 
(Carg\`ese 1978), eds. M.~Levy and S.~Deser, Plenum, New York, 1979.
\bibitem{Wil} T.~J. Wilmore, Riemannian Geometry, (Oxford University Press,
1993).
\bibitem{ls} I.~G. Avramidi, \pl{305}{93}{27}; \jmp{37}{96}{374}.
\bibitem{ZWZ} B. Zumino,Y.-S. Wu and A. Zee, \np{239}{84}{477}.
\bibitem{AGW} L. Alvarez-Gaum\'e and E. Witten, \np{234}{83}{269}.
\bibitem{Ver} J.~A.~M. Vermaseren, Symbolic Manipulation with FORM, (CAN, 
1991).
\bibitem{Wol} S. Wolfram, Mathematica, (Addison-Wesley Pub. Co. , 1991).
%\bibitem{Fu2} S.~A. Fulling, J. Symbolic Comput. {\bf 9} (1990) 73;\ {\it in}
%Proc. of the Third Int. Coll. on Differential Equations, eds. D. Bainov and 
%V. Covachev (VSP Int. Science Publ., 1993).
%\bibitem{GuK} V.~P. Gusynin and V.~V. Kornyak J. Symbolic Comput. {\bf 17} 
%(1994) 283.
\bibitem{MSV} U. M\"uller, C. Schubert and A.~E.~M. van de Ven, work in 
progress.
\bibitem{BoV} J.-P. B\"ornsen and A.~E.~M. van de Ven, work in progress.
\bibitem{FHSS2} D. Fliegner, P. Haberl, M.~G. Schmidt, C. Schubert, The Higher 
Derivative Expansion of the Effective Action by the String-Inspired Method. 
Part II, hep-th/9707189.
\bibitem{Kir} K. Kirsten, The $a_5$ heat kernel coefficient on a manifold 
with boundary, hep-th/9708081.
\end{thebibliography}
\end{document}